\documentclass[12pt]{nature}
\usepackage{times}
\usepackage[utf8]{inputenc}
\usepackage{gensymb}
\usepackage{graphicx}
\usepackage[hidelinks]{hyperref}
\usepackage{multirow}
\usepackage{braket}
\usepackage{amsmath}
\usepackage[T1]{fontenc}
\usepackage{bm}
\usepackage{xcolor}
\usepackage{textcomp}
\usepackage[normalem]{ulem}
\usepackage{times}
\usepackage{chemformula} 

\newcommand{\rev}[1]{#1}

\usepackage[labelfont=bf,labelsep=period]{caption}
\usepackage{float}
\usepackage{newfloat}

\DeclareFloatingEnvironment[
    fileext=lofext,
    listname={List of Extended Data Figures},
    name={Extended Data Fig.}
]{extendedfigure}

\newcommand{\angstrom}{\text{\normalfont\AA}}

\title{Controlling Intertwined Electronic Orders in FeSe with Exfoliation}

\author{Wenyao Liu$^{1}$, Gabriel Natale$^{1}$, Kyung-Mo Kim$^{1}$, Birender Singh$^{1}$, Piyush Sakrikar$^{1}$, Stephen D. Funni$^{2}$, Leo Kondo$^{3}$, Augustin Davignon$^{4}$, Antoine de Lagrave$^{4}$, Kota Ishihara$^{3}$, Michael Geiwitz$^{1}$, Maia G. Vergniory$^{4,5,6}$,
Takasada Shibauchi$^{3}$, Jun Sung Kim$^{7,8}$, Micha{\l} Papaj$^{9}$, Judy J. Cha$^{2}$, Kenneth S. Burch$^{1\dagger}$}
\begin{document}

\maketitle
\begin{affiliations}
\item{Department of Physics, Boston College, Chestnut Hill, MA 02467, USA}
\item{Department of Materials Science and Engineering, Cornell University, Ithaca, USA}
\item{Department of Advanced Materials Science, University of Tokyo, Chiba, Japan}
\item{D\'epartement de physique et Institut quantique, Universit\'e de Sherbrooke, Sherbrooke J1K 2R1 QC, Canada}
\item{Donostia International Physics Center, P. Manuel de Lardizabal 4, 20018 Donostia-San Sebastian, Spain}
\item{Regroupement Qu\'eb\'ecois sur les Mat\'eriaux de Pointe (RQMP), Quebec H3T 3J7, Canada}
\item{Department of Physics, Pohang University of Science and Technology, Pohang 37673, Republic of Korea}
\item{Center for Artificial Low Dimensional Electronic Systems, Institute for Basic Science (IBS), Pohang 37673, Republic of Korea}
\item{Department of Physics and Texas Center for Superconductivity at the University of Houston (TcSUH), Houston, TX 77204, USA}
\\

{$^{\dagger}$To whom correspondence should be addressed; E-mail: ks.burch@bc.edu}
\end{affiliations}

\baselineskip24pt

\begin{abstract}
    Controlling intertwined electronic orders in two-dimensional superconductors offers an effective route to answering fundamental questions and engineering new quantum devices. However, tuning the balance between competing orders typically requires complex chemistry, strain, or interface engineering. Here, we show that a pristine alternative is the dimensional reduction of the unconventional superconductor FeSe. \rev{Exfoliation suppresses the bulk electronic nematic response and switches the superconducting symmetry from bulk $s$-wave to $d$-wave-dominant. Transport, electron microscopy, and Raman spectroscopy establish the substantial weakening of nematic order in thin flakes.} To probe superconductivity, we perform angle-dependent Andreev reflection spectroscopy on pristine crystal edges. As the junction's orientation is rotated, the spectra evolve from zero-energy bound states to coherence peaks. The injection angle, field, and temperature dependence, along with theoretical modeling, confirm that exfoliation switches the superconducting symmetry. Our results suggest a versatile superconducting platform for engineering quantum orders and provide fresh insights into the underlying pairing mechanisms. 
\end{abstract}

\section*{Introduction}  
Exfoliated materials have provided a rich platform for exploring exotic quantum phenomena emerging in low dimensions.\cite{castro2009electronic,tang2017quantum,deng2020quantum,kang2024evidence} While exfoliation began as a tool to access the emergent Dirac quasi-particles in graphene flakes,\cite{castro2009electronic} it has since revealed numerous exotic states of matter.\cite{tang2017quantum,deng2020quantum,kang2024evidence} In several van der Waals materials, exfoliation to the thin limit assists in producing new electronic orders or tuning existing ones, such as transforming a ferromagnet to a superconductor in rhombohedral graphene\cite{zhou2021superconductivity,choi2025superconductivity}, or modifying the balance between charge-density-wave and electron-electron pairing in NbSe$_2$\cite{Xi.2015}, TaS$_2$\cite{kvashnin2020coexistence}, and Kagome superconductors\cite{song2023anomalous}. In these latter systems where superconductivity is intertwined with, and potentially mediated by, electronic orders, exfoliation offers an alternative route to tune the competition without the complications associated with pressure, doping, or thin-film platforms (e.g., strain, charge transfer).\cite{Ahn2021Electrostatic,yu2019high,xie2021fragile} As such, exfoliated superconductors provide a promising avenue for uncovering the underlying pairing mechanisms and realizing new collective states. However, despite reports of new magnetic states\cite{burch2018magnetism} or modified band structures\cite{liu2015electronic,mak2016photonics}, a change in the orbital order or the superconducting symmetry remains rarely observed and typically requires complex heterostructures.\cite{hao2021electric,oh2021evidence,yu2019high,jindal2023coupled,zhao2023time}

In this context, iron-based superconductors (FeSCs) represent an ideal 2D material platform for exploring novel electronic orders. Characterized by strong electronic correlations and intertwined orders among the spin, and nematic degrees of freedom,\cite{kreisel2020remarkable,fernandes2022iron} FeSCs exhibit a remarkable sensitivity of their superconducting properties to structural and electronic changes. Among them, iron selenide (FeSe) stands out due to its simple crystal structure and exceptional tunability.\cite{kreisel2020remarkable} Bulk FeSe hosts electronic nematic order appearing below the transition temperature ($T_{s} \approx 90~K$) that gives way to anisotropic superconductivity below the transition temperature ($T_c\approx 8~K$).\cite{baek2015orbital,yamakawa2016nematicity,liu2018orbital,sprau2017discovery} Interestingly, both ordering temperatures can be significantly modified through chemical substitution\cite{kreisel2020remarkable} or interfacial coupling, as exemplified by the dramatically enhanced $T_{c}$ in monolayer FeSe on SrTiO$_3$.\cite{he2013phase} Substantial attention has also focused on another member of the FeSe family, Fe(Te,Se), where nematic order is completely suppressed, $T_c$ is enhanced, and a topological normal state emerges.\cite{fernandes2022iron} Interestingly, recent studies revealed that exfoliation does not change the $T_c$ of Fe(Te,Se)\cite{tang2019quasi,gray2019evidence}, but can induce a lattice expansion and symmetry breaking at the surface that give rise to a pair-density–modulated superconducting state\cite{kong2025cooper}. 

Inspired by this, we focus on the impact of exfoliation on pristine FeSe. Due to its strong air sensitivity, previous studies on thin flakes have been limited, even though they suggest a weakened nematic transition temperature upon thinning the material.\cite{farrar2020suppression,zhu2021evolution,xie2021fragile} Indeed, transport experiments only reported the decrease of $T_c$ and $T_s$; however, direct probes of the superconducting state, nematic order, and fluctuations are lacking. Here, we use new protocols to enable systematic investigations of exfoliated FeSe over a wide thickness range (15–350 nm; see Supplementary Fig.~S1) through transport, electronic Raman, and Andreev reflection spectroscopies. The schematic of our experimental setup is shown in Fig.~\ref{fig:Exfol}a. These advances are made possible by our inert-atmosphere exfoliation and device fabrication processes, and vacuum suitcase transfer, ensuring the intrinsic quality of FeSe flakes and devices are preserved (see Methods).\cite{gray2020cleanroom} Combined electronic magneto-transport measurements suggest that exfoliation does not substantially alter the Fermi-surface topology. Notably, during this normal-state evolution, \rev{Raman spectroscopy shows that the characteristic bulk-like nematic response is strongly weakened, and that no corresponding transition-like anomaly is resolved in the thin flakes.} Most dramatically, Andreev spectroscopy reveals that the exfoliation tunes the superconducting pairing symmetry of FeSe from nodeless $s$-wave to sign-changing nodal order (as illustrated in Fig.~\ref{fig:Exfol}a) — a phenomenon long pursued through chemical or interface engineering.\cite{kreisel2020remarkable,ishida2022pure} Taken together, these results highlight exfoliation as a clean and versatile route to tune the intertwined nematic and superconducting states in FeSe, unveiling an unprecedented degree of quantum control in a single material system.

\section*{Electrical Transport}
\begin{figure}[H]
    \centering
    \includegraphics[width=0.85\linewidth]{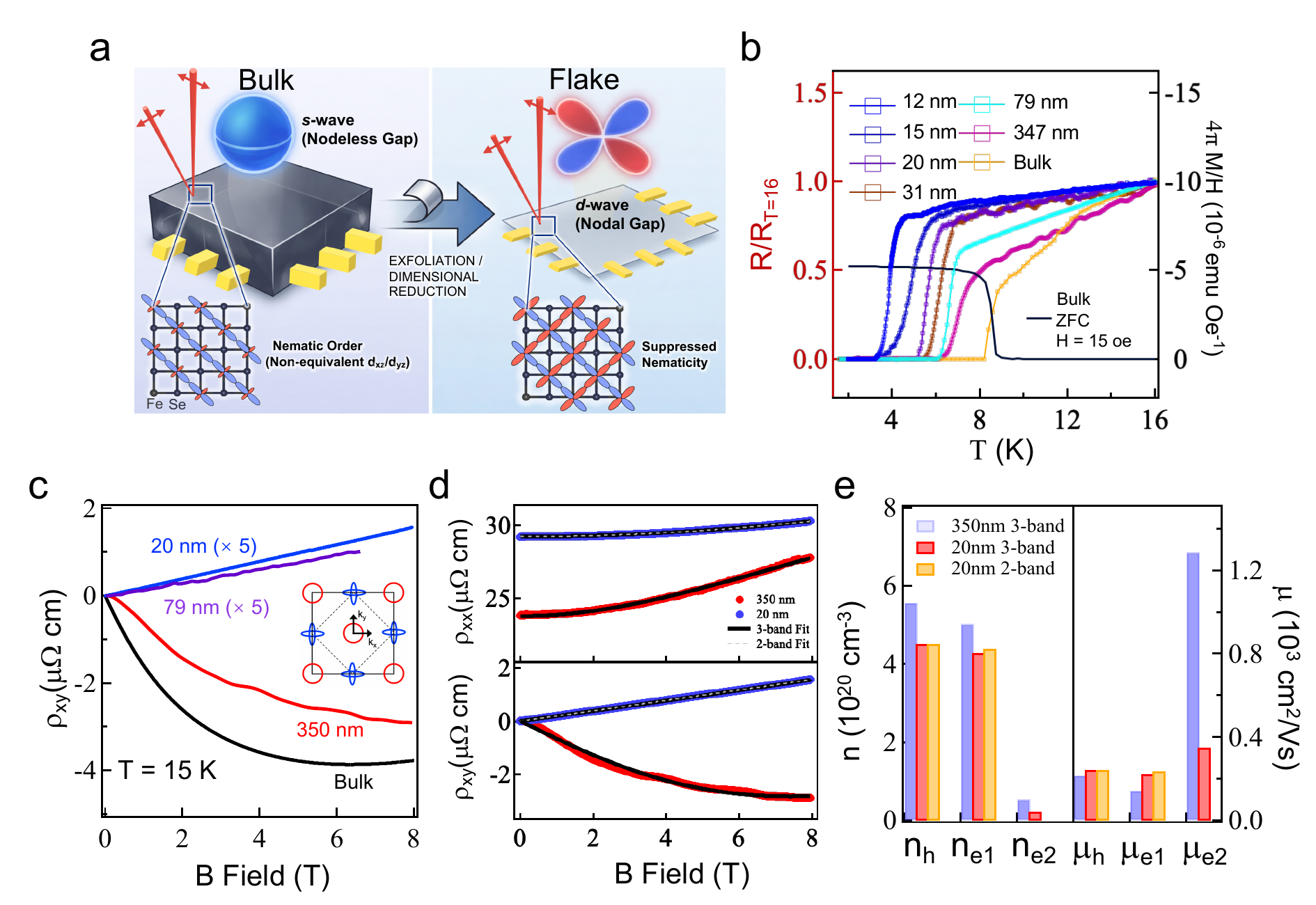}
    \caption{\label{fig:Exfol} \textbf{Exfoliated FeSe to thin flakes.} \textbf{a,} Switching intertwined order in FeSe through exfoliation. The devices on exfoliated and bulk FeSe crystal are used for ERS and transport measurements. \textbf{b,} Temperature dependence of the normalized resistance \textit{R}(\textit{T})/\textit{R}(16K) for a bulk crystal and thin flakes with different thickness, compared with the magnetic susceptibility of bulk crystal (black curve). \textit{R}(\textit{T}) of bulk crystal is extracted from the previous work\cite{watson2015dichotomy}, compared with our magnetic susceptibility. \textbf{c,} Hall measurement on different thickness FeSe at T = 15 K. The bulk curve is from the previous work\cite{watson2015dichotomy}. Inset: Hole-like (red) and electron-like (blue) Fermi surfaces of FeSe single crystals. The solid and dashed squares indicate the 1-Fe and 2-Fe Brillouin zone, respectively. \textbf{d,} Simultaneous fit of the $\rho_{xx}$ and $\rho_{xy}$ on 350 nm- and 20 nm- flakes, using a two-band or three-band model\cite{watson2015dichotomy}. \textbf{e,} Carrier density $n$ and mobility $\mu$ extracted from fits in (\textbf{d}).}
\end{figure}

We first examined the change of $T_c$, using resistance versus temperature measurements ($R$–$T$) (Fig.~\ref{fig:Exfol}b). Consistent with previous reports, $T_c$ decreases from $\sim8$ K in bulk to $\sim4$ K near 15 nm.\cite{farrar2020suppression,zhu2021evolution} To further probe potential changes in Fermi pockets, some studies examined the normal-state (T $ = 15\, K  > $ $T_c$) Hall resistivity $\rho_{xy}(B)$\cite{zhu2021evolution}. As shown in Fig.~\ref{fig:Exfol}c, we also found the thick ($\sim$ 350 nm) flake shows a nonlinear, negative $\rho_{xy}(B)$, while thinner flakes exhibit a nearly linear, positive $\rho_{xy}(B)$. Since the Hall response is associated with the carrier densities (see Method)\cite{watson2015dichotomy}, this could result from changes of the Fermi-surface topology in thin, exfoliated FeSe. However, without doping or charge transfer that occurs from the substrate in thin films\cite{kreisel2020remarkable}, it is unlikely that the relative population of electron- and hole-pockets has largely changed. 

Alternatively, a similar linear-to-nonlinear transformation of $\rho_{xy}(B)$ upon cooling in bulk FeSe is associated with the nematic order induced changes in mobility\cite{watson2015dichotomy}. Specifically, the $\rho_{xy}(B)$ is B-linear when $T$ > $T_s$, due to the similar mobilities and densities of the electron and hole pockets (see schematics in Fig.~\ref{fig:Exfol}c inset and Methods). Upon cooling below $T_s$, the band reconstruction generated by the nematic order turns an electron pocket into high-mobility Dirac-like fermions, producing the non-linear B-dependence $\rho_{xy}$.\cite{watson2015dichotomy} As shown in Fig.~\ref{fig:Exfol}d,e, the longitudinal $\rho_{xx}(B)$ and Hall resistivity $\rho_{xy}(B)$ of thick and thin flakes are simultaneously fit to extract the carrier density $n$ and mobility $\mu$. \rev{Here, the 20 nm flake can be described by a compensated two-band ($n_{h}$ and $n_{e1}$) or three-band ($n_{h}$, $n_{e1}$ and $n_{e2}$) model with well-constrained parameters. The carrier density ($n_{e1}$) in the three-band fit of the 350 nm flake is consistent with a range of $(5$-$10)\times10^{20}~\mathrm{cm^{-3}}$, and thus cannot provide a precise quantitative comparison between thick and thin flakes. However, reproducing the nonlinear Hall response of the 350 nm flake requires a three-band model with an additional high-mobility electron component ($\mu_{e2}$).\cite{watson2015dichotomy} This qualitative distinction suggests that the bulk-like nematic reconstruction of the low-energy electronic structure is strongly suppressed in thin flakes.}

\section*{Nematic Evolution}
\begin{figure}[t]
    \centering
    \includegraphics[width=1\linewidth]{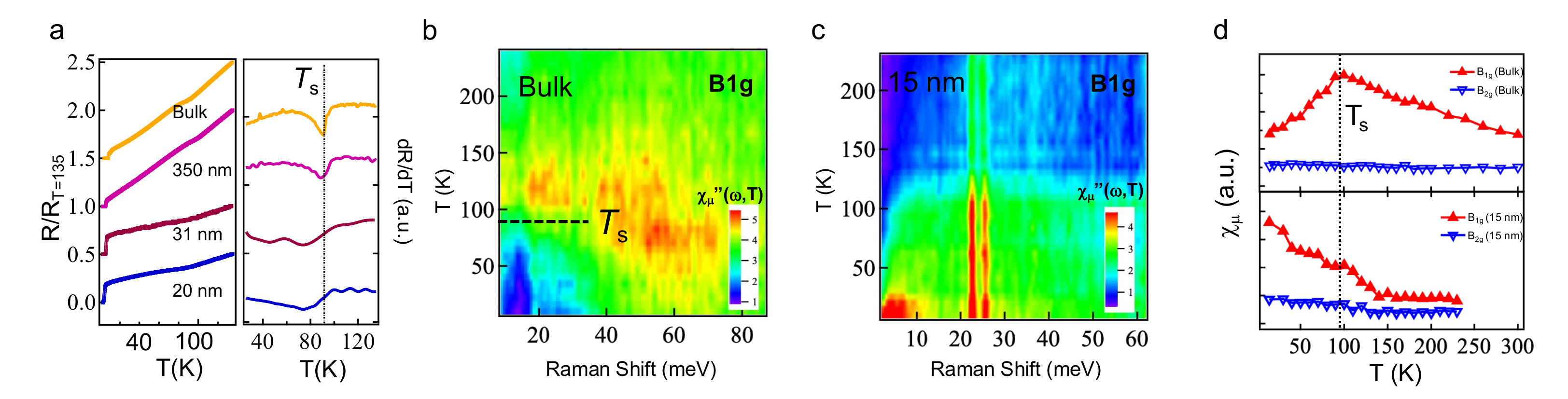}
     \caption{\label{fig:Nematicity} \textbf{Nematicity in bulk and thin FeSe} \textbf{a,} Temperature dependence of the normalized resistance R(T)/R(135K) and corresponding derivative dR/dT for a bulk crystal\cite{watson2015dichotomy} and exfoliated flakes. Curves are shifted for clarity. 
     \textbf{b,} Temperature--Raman-shift map of the bulk $B_{1g}$ Raman response, $\chi_{B_{1g}}''(\omega,T)$. The horizontal dashed line marks $T_s$. \textbf{c,} Corresponding $B_{1g}$ Raman-response map for a 15-nm FeSe flake. The bulk-like nonmonotonic enhancement near $T_s$ is not resolved. \textbf{d,} Finite-window dynamic susceptibilities in the $B_{1g}$ and $B_{2g}$ channels. Top: bulk FeSe. Bottom: 15-nm FeSe. The upper integration cutoff is 50~meV.
     }
\end{figure}

Inspired by the change in the Hall response, we next examine the evolution of the bulk-like nematic state. In bulk FeSe, nematic order is directly probed by NMR experiments below 90 K.\cite{baek2015orbital} Here, Fig.~\ref{fig:Nematicity}a displays $R(T)$ and its derivative for bulk and exfoliated FeSe. The bulk crystal and 350\,nm flake exhibit a resistive kink near $T_s \approx$ 93 K, associated with the tetragonal-to-orthorhombic transition and the onset of nematic order.\cite{watson2015dichotomy,farrar2020suppression} In thin flakes (31 nm and 20 nm), this kink is replaced by a weaker and broader inflection at a lower temperature ($\sim$ 75 K), similar to previous work\cite{zhu2021evolution}, and consistent with a strong suppression of the nematic instability. 

\rev{To further probe this evolution, we performed temperature-dependent, symmetry-resolved electronic Raman scattering (Fig.~\ref{fig:Nematicity}b-d and Extended Data Figs.~\ref{ExtenedFig:BulkRaman} and \ref{ExtenedFig:FlakeRaman}). For micron-scale exfoliated samples, symmetry-resolved Raman scattering is particularly well suited because it is contactless, can be focused onto individual flakes, and probes the electronic nematic susceptibility without externally manipulating the sample. In addition, electronic Raman scattering is well established as a symmetry-resolved probe of nematic fluctuations in Fe-based superconductors.\cite{gallais2013observation,gallais2016charge} In FeSe specifically, the low-energy Raman response has further been used to track the weakening of nematicity under sulfur substitution and pressure.\cite{massat2016charge,zhang2021quadrupolar,chibani2021lattice, massat2018collapse} In the one-Fe notation, the nematic electronic response is observed in the $B_{1g}$ channel, whereas the bulk $B_{2g}$ response provides a non-nematic symmetry reference.\cite{massat2016charge,zhang2021quadrupolar,chibani2021lattice} Previous bulk measurements established two characteristic components of the $B_{1g}$ response: a low-energy quasi-elastic peak (QEP) and a broad higher-energy electronic response centered near 40-60 meV.\cite{massat2016charge,zhang2021quadrupolar,chibani2021lattice} The QEP has the strongest critical temperature dependence, increasing on cooling toward $T_s$ and diminishing below the transition. The higher-energy component is considerably less critical across $T_s$, although temperature-dependent reconstruction of this continuum below $T_s$ has also been associated with the nematic state.\cite{zhang2021quadrupolar,chibani2021lattice} This evolution can be visualized directly in temperature--Raman-shift maps (< 50 meV) and quantified through the corresponding low-energy dynamic susceptibility.\cite{massat2016charge,zhang2021quadrupolar} The same Raman phenomenology has been used to follow the evolution of nematicity under external tuning. In FeSe$_{1-x}$S$_x$ and pressure-tuned FeSe, the low-energy QEP evolves systematically as $T_s$ is suppressed.\cite{zhang2021quadrupolar,chibani2021lattice,massat2018collapse}}

\rev{Our bulk measurements reproduce these established signatures. The $B_{1g}$ Raman map in Fig.~\ref{fig:Nematicity}b shows a pronounced enhancement of low-energy spectral weight on approaching $T_s$, followed by its suppression upon cooling further into the nematic phase (T < $T_s$). The corresponding wide-energy spectra are shown in Extended Data Fig.~\ref{ExtenedFig:BulkRaman}a,b and display both the low-energy QEP and the broad higher-energy $B_{1g}$ response in the 40--60-meV range, consistent with previous bulk Raman measurements.\cite{massat2016charge,chibani2021lattice} We note that this energy range is comparable to the characteristic energy scale of the nematic electronic reconstruction observed by ARPES, where the $d_{xz}$--$d_{yz}$ band splitting reaches approximately 50 meV.\cite{yi2019nematic} In contrast, the bulk $B_{2g}$ spectra and color map show no corresponding critical enhancement near $T_s$ (Extended Data Fig.~\ref{ExtenedFig:BulkRaman}c,d). Consistent with this low-energy spectral evolution, using the Kramers--Kronig procedure (see Methods), the $B_{1g}$ dynamic susceptibility, $\chi_{B_{1g}}(0,T)$, for this finite window in Fig.~\ref{fig:Nematicity}d develops a pronounced maximum near $T_s$, whereas the corresponding $\chi_{B_{2g}}(0,T)$ remains smooth.}

\rev{We next turn to the thin flakes. Figure~\ref{fig:Nematicity}c shows the $B_{1g}$ Raman map of a 15-nm flake, while the full $B_{1g}$ and $B_{2g}$ spectra of independent 15-nm and 23-nm flakes are shown in Extended Data Fig.~\ref{ExtenedFig:FlakeRaman}a-d. Compared with bulk FeSe, the broad electronic response in the 40--60-meV range becomes substantially less pronounced in the thin-flake spectra, making the phonon near 25 meV more prominent relative to the surrounding electronic response. We note that a comparable reduction of the higher-energy $B_{1g}$ spectral weight has been reported with increasing S substitution in FeSe$_{1-x}$S$_x$, and discussed in terms of both electronic/quadrupolar and local-moment spin excitations.\cite{chibani2021lattice,lazarevic2022evolution} Although the microscopic interpretation of this component is not unique, the similarity indicates that its reduced prominence is consistent with an intrinsic evolution of the FeSe electronic Raman response. Therefore, we regard this high-energy spectral redistribution as supporting evidence for electronic reconstruction upon thinning, rather than as the primary measure of the nematic transition. Instead,  we employ a more direct Raman signature: the low-energy temperature evolution. Unlike bulk FeSe, where the $B_{1g}$ response shows a pronounced nonmonotonic enhancement on approaching $T_s$, neither the 15-nm nor the 23-nm flake develops a comparable transition-like redistribution of low-energy Raman spectral weight down to the lowest measured temperatures of 14~K (Fig.~\ref{fig:Nematicity}c and Extended Data Fig.~\ref{ExtenedFig:FlakeRaman}a,b).}

\rev{Furthermore, the same distinction is evident in the finite-window partial dynamic susceptibilities (Fig.~\ref{fig:Nematicity}d and Extended Data Fig.~\ref{ExtenedFig:FlakeRaman}e). Whereas the bulk $B_{1g}$ response develops a pronounced nonmonotonic maximum near $T_s$, the nominal $B_{1g}$ responses of both thin flakes evolve smoothly and show no corresponding transition-like anomaly. This conclusion remains unchanged across several integration windows (Supplementary Fig.~S11). In the thin flakes, the nominal $B_{2g}$ responses also exhibit a modest, smooth temperature dependence (Extended Data Fig.~\ref{ExtenedFig:FlakeRaman}f). One plausible contribution to this smooth background is the known temperature-dependent Raman response of heavily doped Si, as illustrated by the representative substrate measurement in Extended Data Fig.~\ref{ExtenedFig:FlakeRaman}g.\cite{cerdeira1973effect,hart1970temperature} In addition, to minimize artificial laser-heating effects, we used sufficiently low excitation powers and independently evaluated it from the Stokes-to-anti-Stokes intensity ratio (Supplementary Information Section~VIII).\cite{wang2020range}}

\rev{Taken together, the Raman measurements show that the characteristic bulk-like nematic transition is strongly weakened and no longer resolved in either thin flake down to 14 K. This does not exclude residual nematic correlations in the exfoliated samples. Rather, together with the suppression of the resistive anomaly and the disappearance of the high-mobility Hall component, these results indicate a pronounced weakening of the bulk-like nematic reconstruction in thin FeSe.}

\section*{Superconducting Order Change}
\begin{figure}[t]
    \centering
    \includegraphics[width=0.8\linewidth]{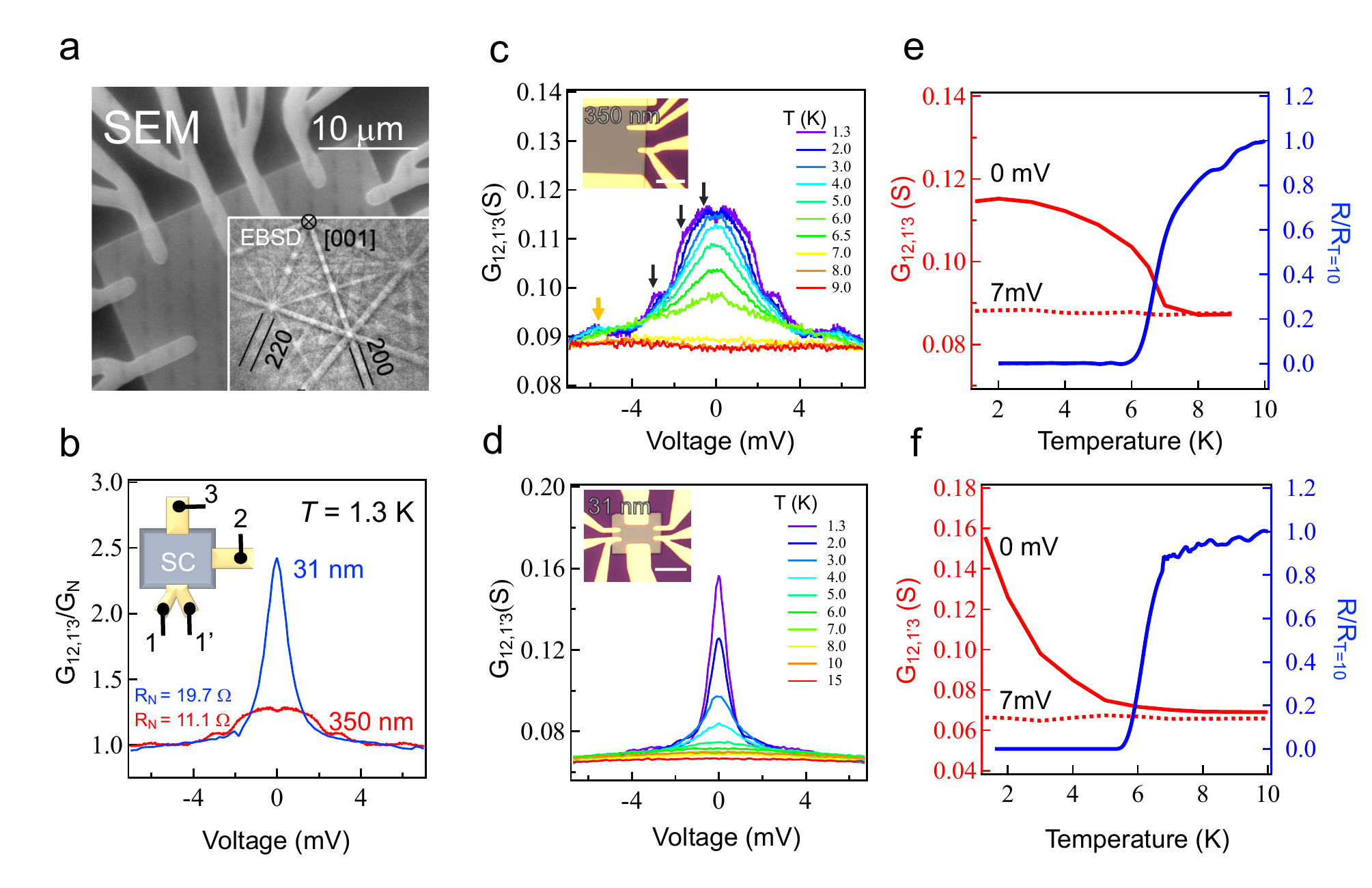}
    \caption{\label{fig:AR_bulkthin} \textbf{Superconducting spectra changing upon exfoliation} \textbf{a,} SEM image of the FeSe device, Inset: the back-scattering diffraction pattern show that the electrodes contact the naturally-cleaved edges along crystalline [100]-direction. \textbf{b,} Differential conductance $G_{12,1'3}$ of edge contacts on 350 nm and 31 nm thickness devices probed along $\alpha = 0$ measured at 1.3 K. Inset: the configuration of the measurement. Here, $G_{ij,kl} = dI_{ij}/dV_{kl}$, where i,j represent the current source (I+) and drain (I-),  k,l represent the voltage leads (V+ and V-). All conductance curves are normalized by the normal-state conductance ($G_{N} = 1 / R_{N}$). \textbf{c,} Temperature dependence of the differential conductance for the 350 nm thick device. Inset: Device image with 5 $\mu$m scale bar. \textbf{d,} Temperature dependence of the differential conductance for the 31 nm thick device. \textbf{e,} and \textbf{f,} temperature evolution of conductance at 0 mV and 7 mV extracted from (\textbf{c}) and (\textbf{d}), and normalized resistance \textit{R}(\textit{T})/\textit{R}(10K) probed on the same device using a four-terminal configuration.
    }
\end{figure}

The dramatic changes in nematicity and the superconducting $T_c$ indicate that exfoliation may also alter the superconducting order parameter (sOP) of FeSe. Hence, we turned to angle-dependent Andreev reflection spectroscopy (ARS), a powerful technique to explore sOP by detecting the gap size and the presence of nodal or sign-changing phase.\cite{kashiwaya2000tunnelling,daghero2011directional} ARS experiments on 2D materials remain rare, given the challenges of preparing sharp edges along specific crystallographic orientations, while simultaneously achieving sufficiently low-barrier and small-size contacts.\cite{efetov2016specular} Here, we exploit the naturally cleaved edges of tetragonal FeSe crystals, where submicron electrodes are fabricated carefully within an argon atmosphere (see Methods).\cite{gray2020cleanroom} For reliable comparison between samples, all contacts are precisely aligned, verified by scanning electron microscopy (SEM; Fig.~\ref{fig:AR_bulkthin}a and Supplementary Fig.~S2b,d). Electron backscatter diffraction (EBSD; Fig.~\ref{fig:AR_bulkthin}a inset and Supplementary Fig.~S2c,e) confirmed that the crystal orientations of flake edges are primarily normal to [100]/[010]. Notably, our fabrication process (see Methods) yields twin-branch ARS contacts that selectively probe the flake edges. This approach eliminates common artifacts in ARS, such as strain-induced distortions and large resistance backgrounds from metal tips/leads (see Supplementary Information Section~I).\cite{daghero2011directional,chen2022tip} The spectroscopic (i.e., ballistic) nature of these contacts, and the injection of current normal to the edge, are confirmed by the effective contact size (see Supplementary Information Section~I) and, as discussed later, by the magnetic-field dependence of ARS spectra.

Starting with bulk FeSe, we note the established multi-band, highly anisotropic, but dominant $s$-wave order with the gap minima/maxima along [110]/[1$\overline{1}$0] aligned to the nematic order\cite{liu2018orbital,sprau2017discovery}. Here, due to the presence of nematic twin domains in bulk FeSe (the size can be narrow as tens of nm)\cite{watashige2015evidence}, ARS at [100]/[010] edges should provide a representative average over this anisotropic gap, with multi-coherence-peak features\cite{tortello2010multigap,daghero2011directional}. As shown in Fig.~\ref{fig:AR_bulkthin}b, differential conductance was measured in a three-terminal configuration utilizing twin-branch contacts ($G_{12,1^{'}3} = dI_{12}/dV_{1^{'}3}$; Fig.~\ref{fig:AR_bulkthin}b inset) on the quasi-bulk (350 nm) flake (inset of Fig.~\ref{fig:AR_bulkthin}c). The AR spectra exhibit symmetric sub-gap maxima with $G/G_N \ll 2$ (Fig.~\ref{fig:AR_bulkthin}b, c), are consistent with a multi-band, nodeless superconducting state\cite{daghero2011directional,bourgeois2016thermal}. Indeed, coherence-peak features (subgap maxima marked by black arrows in Fig.~\ref{fig:AR_bulkthin}c) reveal a multi-gap structure. In addition, two shoulder-like anomalies around 6 mV are observed (orange arrows), probably induced by strong electron-boson coupling, consistent with those reported in FeTe$_{1-x}$Se$_x$ and Fe-pnictide superconductors \cite{tortello2010multigap,daghero2011directional} at the characteristic energy of the spin resonance\cite{chen2019anisotropic}.

We note that our ARS spectra clearly capture the multi-gap features in bulk FeSe, not observed in traditional point-contact experiments\cite{naidyuk2017superconducting}, but are in good agreement with STM\cite{sprau2017discovery}. Furthermore, the temperature dependence seen in Fig.~\ref{fig:AR_bulkthin}c is consistent with thermal smearing, while all Andreev features disappear at $T_{c}$. This is best illustrated in Fig.~\ref{fig:AR_bulkthin}e, which plots the temperature dependence of the device resistance and $G_{12,1^{'}3}$ at zero bias (i.e., the Andreev contribution) and at 7 meV (i.e., the normal state). Crucially, the low- and high-bias features meet at $T^{A}_{c} \approx$ 7 K, consistent with the $T_{c}$ determined from R-T curves, confirming the absence of local heating and strain effects\cite{chen2022tip}. Meanwhile, the high-bias conductance is nearly temperature-independent. Thus, we demonstrate that our fabricated contacts are well within the Sharvin (i.e., ballistic) regime (see Supplementary Information section I) \cite{daghero2011directional}, and thus offer highly reliable ARS.

Having established the ability to probe ARS in thick flakes, we turned to thin flakes with contacts normal to the same crystal directions ([100]/[010]). As shown in Fig.~\ref{fig:AR_bulkthin}b, the 31 nm flake exhibited an AR spectrum radically different from the bulk, despite having a similar $T_{c}$. The coherence peak features are strongly suppressed, and instead a pronounced zero-bias conductance peak (ZBCP) with $G/G_N > 2$ is observed. Notably, this distinct feature was observed in five FeSe flakes with thicknesses ranging from 20 to 79 nm (see Supplementary Fig.~S3b, c). This ZBCP is similar to those observed in cuprates and heavy-fermion superconductors with the current applied along the nodal direction, a hallmark of a zero-energy Andreev bound state (ABS).\cite{kashiwaya2000tunnelling,daghero2011directional,zareapour2017andreev} Crucially, the contacts exhibit several features consistent with intrinsic rather than extrinsic factors. In particular, the temperature dependence shown in Fig.~\ref{fig:AR_bulkthin}d matches the expected suppression of an ABS. Furthermore, as shown in Fig.~\ref{fig:AR_bulkthin}f, the normal-state (high-bias) conductance ($G_N$) remains temperature-independent, while the ZBCP reaches the normal state at $T_{c}$, as seen in R(T). Lastly, from the measured $G_N$, we estimated that the effective contact sizes of all flakes are on the nanometer scale  (see Supplementary Fig.~S3a-d and Supplementary Information Section~I). This is consistent with the flake thickness and, more importantly, supports ballistic contacts with minimal Maxwell or heating effects.\cite{daghero2011directional}  

Next, we turn to analyze the ARS using the established 2D Blonder--Tinkham--Klapwijk (2D-BTK) framework (Extended Data Fig.~\ref{ExtenedFig:SymmetryARSModel}, and Supplementary Information section II)\cite{kashiwaya2000tunnelling,daghero2011directional}. In this model, a sign-preserving sOP (e.g., extended $s$-wave, Extended Data Fig.~\ref{ExtenedFig:SymmetryARSModel}a) can produce a small zero-bias bump when it carries nodes ($\Delta_{minimum} = 0$, Extended Data Fig.~\ref{ExtenedFig:SymmetryARSModel}b). However, this occurs only when temperature, lifetime broadening ($\Gamma$), and the dimensionless barrier (Z) are all zero. Nonetheless, the normalized differential conductance is always less than two ($G/G_N < 2$). In realistic devices where T, $\Gamma$, and Z are all greater than 0, this bump is replaced by coherence peaks whose energy is weakly dependent on the current-injection angle $\alpha$. Indeed, in quasi-bulk FeSe, the ARS displays clear coherence peaks with amplitudes $G/G_N \leq 1.3$ (red curve in Fig.~\ref{fig:AR_bulkthin}b). In contrast, the thin flakes ARS (e.g., the blue curve in Fig.~\ref{fig:AR_bulkthin}b) with ZBCP ($G/G_N \approx 2.3$) and without coherence peaks are consistent with a sign-changing nodal sOP (e.g., $d$-wave superconductor, Extended Data Fig.~\ref{ExtenedFig:SymmetryARSModel}c). Specifically, when current is injected parallel to the nodal direction ($\alpha$ = 0, $Z, T,\Gamma >0$), a ZBCP with amplitude $G/G_N > 2$ arises from an ABS due to constructive interference between electron-like and hole-like quasiparticles generated by the sOP with opposite signs (Extended Data Fig.~\ref{ExtenedFig:SymmetryARSModel}d).

\begin{figure}[t]
    \centering
    \includegraphics[width=0.55\linewidth]{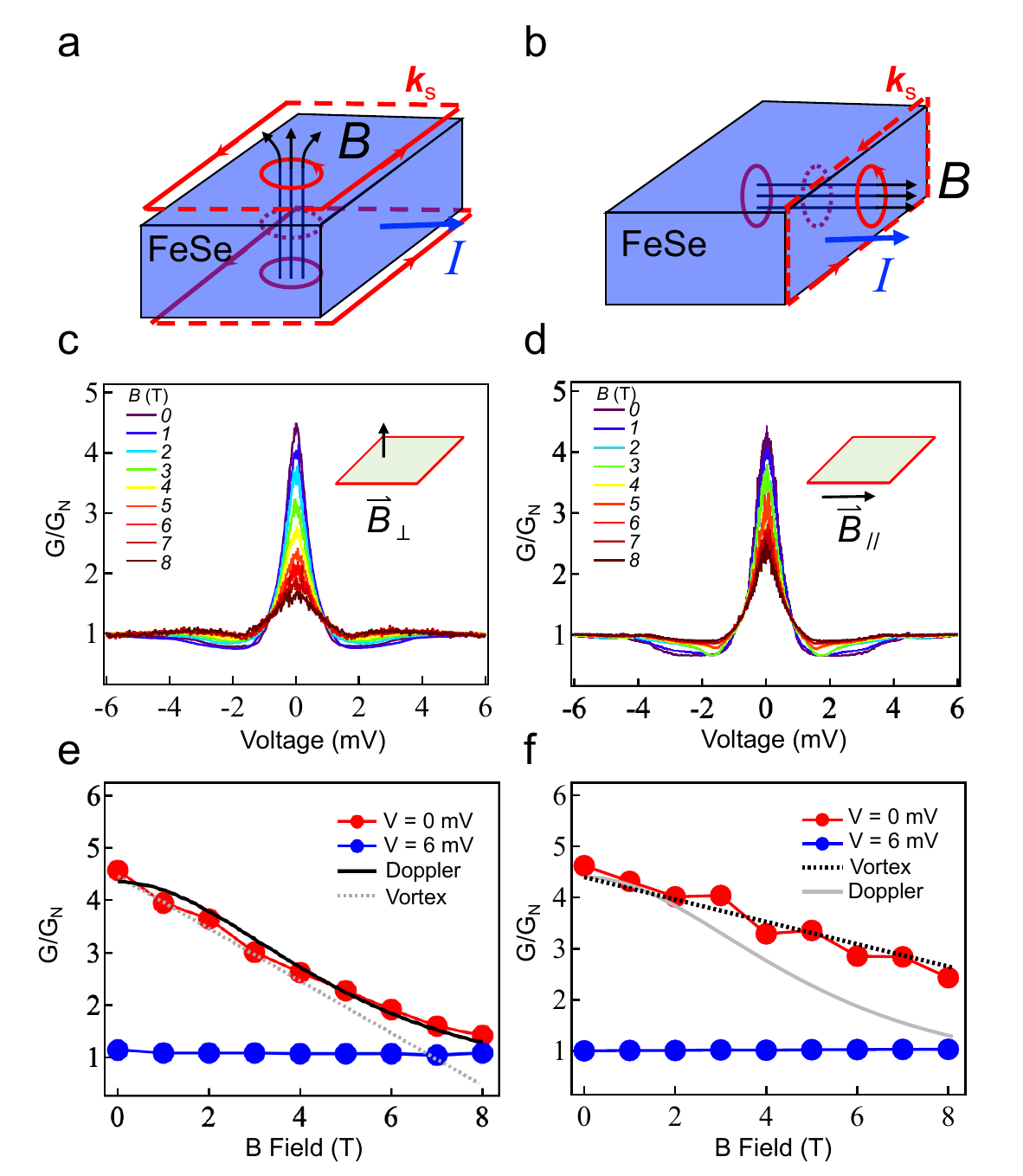}
    \caption{\label{fig:AR_Bfield} \textbf{Response to an external magnetic field.} Schematics of applying external magnetic field $B$ along \textbf{a}, out-of-plane, and \textbf{b}, parallel directions with respect to the tunneling current $I$. Red arrow lines indicate the SC order-parameter phase gradient, $k_{s}$. Black arrow lines represent magnetic flux. \textbf{c,} and \textbf{d,} The normalized conductance curves at 1.4 K for an edge contact probed along $\alpha$ = 0 with \textbf{c,} $\vec{B}_{\perp}$, and \textbf{d,} $\vec{B}_{||}$. \textbf{e,} and \textbf{f,} Corresponding zero-bias and 6 mV-bias conductance versus magnetic field from (\textbf{c}) and (\textbf{d}) with the curves fit (back) and simulation (gray) via Doppler model (solid line) and vortex-state smearing effects model (dashed line). The Doppler shift is set as $\Delta E \approx 0.125~B$ in both (\textbf{e}) and (\textbf{f}). The vortex radius $r_V$ $\approx$ 10.5 nm for the simulation in (\textbf{e}), and 6.49 nm extracted by fit in (\textbf{f}).} 
\end{figure}

In this 2D-BTK analysis, one assumes that the in-plane injection current dominates the c-axis contribution to ARS results. To test this, we exploited the expected dependence of the Andreev conductance on the orientation of the magnetic field ($\vec{B}$) relative to the injected current ($\vec{I}$).\cite{zareapour2017andreev,rohlfing2009doppler} Specifically, as shown in Fig.~\ref{fig:AR_Bfield}a, when $\vec{B} \perp \vec{I}$, a Doppler shift arising from tunneling into the screening supercurrent modifies the quasiparticle energies, producing a non-linear and anisotropic field response (see Supplementary Information section V).\cite{zareapour2017andreev,rohlfing2009doppler} In contrast, as seen in Fig.~\ref{fig:AR_Bfield}b, for $\vec{B} \parallel \vec{I}$, vortices smear and suppress the Andreev features, resulting in a nearly linear reduction of conductance with field (see Supplementary Information section V).\cite{rohlfing2009doppler,zareapour2017andreev} Consistent with this, the 1.3 K normalized ARS shown in Fig.~\ref{fig:AR_Bfield}c is rapidly reduced with an out-of-plane B-field. Furthermore, as shown in Fig.~\ref{fig:AR_Bfield}e, the ZBCP suppression is well fit by the Doppler-effect model ($G(V) \to G(V+\Delta V)$, where $\Delta V$ = $D\left|B\right|$) with $D$ = 0.124, rather than the vortex-core smearing (dashed line). Crucially, the conductance above $\pm$4 mV is nearly field-independent, indicating the absence of magnetoresistance or Doppler response when tunneling into the normal state. Next, we measured the ARS curves as a function of in-plane field. As shown in Fig.~\ref{fig:AR_Bfield}d, the ABS suppression is slower than with the c-axis field. Indeed, the suppression of AR conductance matches vortex-core smearing (dashed line in Fig.~\ref{fig:AR_Bfield}f), with vortex radius $\approx 6.5$ nm as reasonably expected for FeSe\cite{farrar2020suppression,putilov2019vortex}. Again, the $G_N$ does not respond to the in-plane applied field. These results confirm that the ARS curves are mainly due to in-plane currents in FeSe flakes.

\section*{Symmetry of the Order Parameter in Thin FeSe}
Having established the current flows normal to the side surface, we now turn to test the dependence on the injection angle to explore the possible symmetry of sOP in thin FeSe. In a sign-changing nodal sOP case (e.g, $d$-wave), upon moving the detection away from the node (either in-plane or along the c-axis\cite{daghero2011directional,kashiwaya2000tunnelling}), the ZBCP should be continuously suppressed. Moreover, the coherence peaks will appear and grow in strength when the angle is moved towards the anti-node. In a mixture case, e.g., $s$+$d$ wave, where a $d$-wave order parameter is mixed with a strong $s$-wave order (that can lift the node, $\Delta_d \le \Delta_s$), the ZBCP will disappear due to the lack of ABS (see Extended Data Fig.~\ref{ExtenedFig:SymmetryARSModel}e).

\begin{figure}[t]
    \centering
    \includegraphics[width=0.5\linewidth]{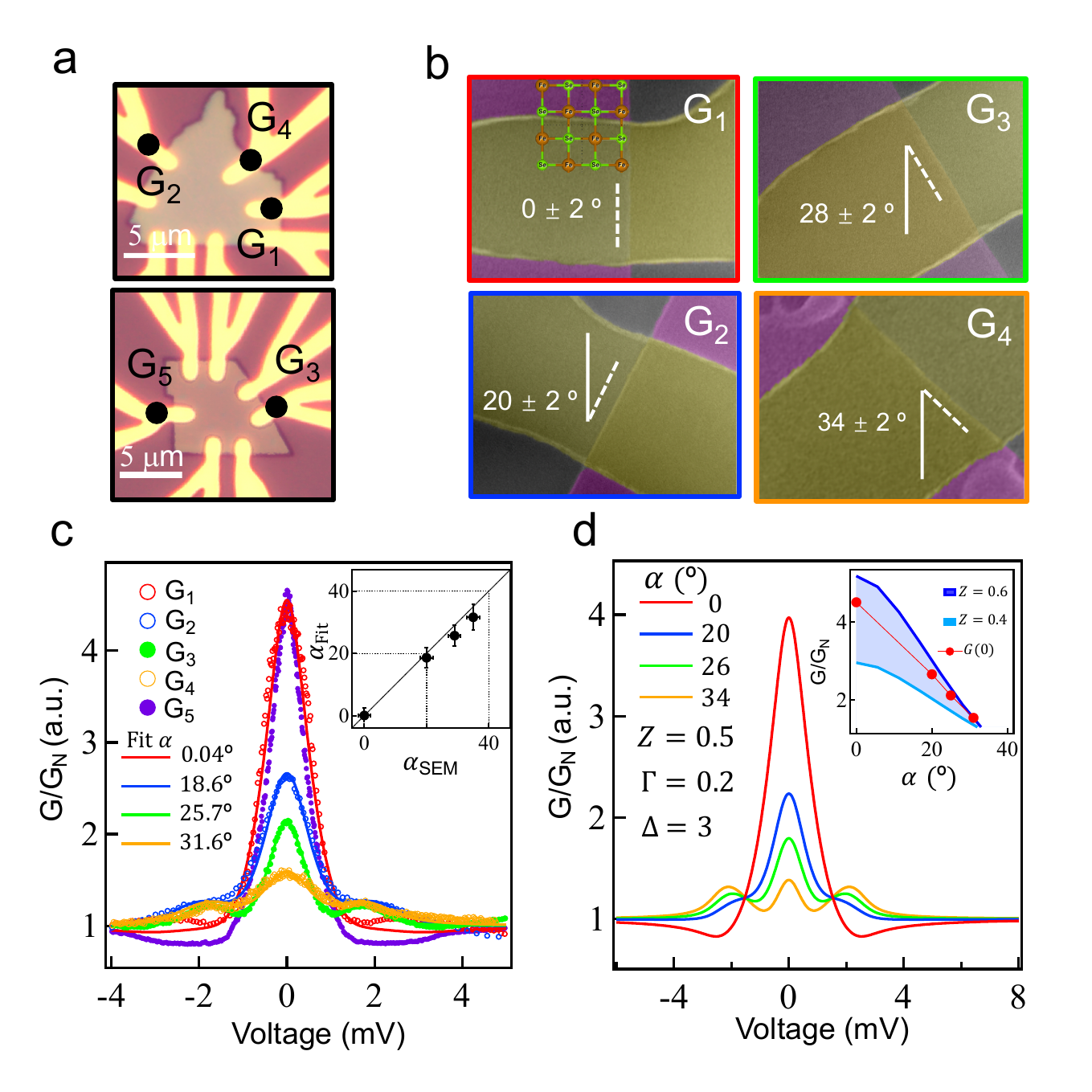}
    \caption{\label{fig:directionalAR} \textbf{Symmetry of sOP in thin flake} \textbf{a,} Optical image of two 20 nm-thickness FeSe devices, with contacts ($G_1$ to $G_5$) on different-orientation edges. \textbf{b,} False-color SEM images of the contact (yellow) on FeSe devices (purple). The angle relative to the edge normal [100] is determined by EBSD. \textbf{c,} The 1.3K, normalized conductance of contact $G_{1}$ to $G_{5}$ (symbols) and the relevant 2D-BTK fit assuming a $s$+$d$-wave gap (line). The parameters of the model are gap amplitude $\Delta_s$, $\Delta_d$, the barrier strength Z, the angle $\alpha$, and a Dynes parameter $\Gamma$ (describing the quasi-particle's lifetime in the contact\cite{daghero2011directional}).  Inset: $\alpha_{Fit}$ is extracted from the BTK fit of $G_{1}$ to $G_{4}$ curves, $\alpha_{SEM}$ is obtained directly from SEM images in (\textbf{b}). \textbf{d,} Theoretical normalized conductance curves at T = 1.3 K calculated with the 2D-BTK model assuming $\alpha$ from 0 to 34 $^{\circ}$ with all other parameters fixed ($Z = 0.5, \Gamma = 0.2, \Delta = 3$). Inset: the normalized conductance at zero-bias voltage versus angle $\alpha$ from 2D-BTK model (lines) for a range of barrier heights and the experimental values (symbols) extracted from curves in (\textbf{c}). 
    }
\end{figure}

Hence, we turned to angle-dependent ARS measurement by employing edges with varying normal vectors. As shown in Fig.~\ref{fig:directionalAR}a, two flakes (thickness $\sim$20 nm) with multiple clean edges provide different contacts for varying electron-injection angles $\alpha$ relative to [100]. To precisely determine $\alpha$, SEM (Fig.~\ref{fig:directionalAR}b) and EBSD (Supplementary Fig.~S2) were implemented on the same devices. As seen in Fig.~\ref{fig:directionalAR}b, the edges under the contacts are nearly perfectly cleaved with $\alpha = 0^\circ, 20^\circ, 26^\circ, 34^\circ$. Fig.~\ref{fig:directionalAR}c shows the normalized conductance $G(V)/G_N$ at $T=1.3\,\mathrm{K}$ for five contacts ($G_1$--$G_5$) measured for these two flakes. As expected for a $d$-wave sOP, AR conductance curves should exhibit a continuous $\alpha$ dependence.\cite{kashiwaya2000tunnelling,daghero2011directional}  At $\alpha = 0^{\circ}$ on both flakes ($G_1$ and $G_5$ in Fig.~\ref{fig:directionalAR}a), a pronounced ZBCP with $G/G_N \approx 4.5$ is observed, while rotating $\alpha$ from 0$^{\circ}$ to 34$^{\circ}$ ($G_1$ to $G_4$), gradually suppresses the ZBCP, and coherence peaks around 2 mV are continuously enhanced. 

In addition, we note that a ZBCP-like feature may also arise from a nodeless, sign-changing sOP—such as a $s\pm$ state—when the inter-band coupling is finely tuned (see Supplementary Information Section III). However, unlike bulk FeSe where the gap minima lie along the Fe–Fe direction, the observation of the strongest ZBCP and the absence of coherence peaks at $\alpha = 0^{\circ}$ identify Fe-Se orientation as the gap nodal direction in thin flakes, directly contradicting the $s\pm$-wave ABS scenario.\cite{liu2018orbital,sprau2017discovery} Moreover, the ZBCP in thin-flake FeSe exhibits a continuous evolution over a wide angular range ($0^\circ\rightarrow 34^\circ$), further ruling out the possible $s\pm$-wave scenario (see Supplementary Information Section III). 

To quantitatively analyze the $\alpha$-dependent ARS curves, we employed the 2D-BTK model by assuming a $s$+$d$-wave gap function ($\Delta(\alpha) = \Delta_s + \Delta_d \cos(2\alpha+\frac{\pi}{2}),
$ where $\Delta_s$ and $\Delta_d$ are amplitudes of $s$-wave and $d$-wave gap) \cite{daghero2011directional,kashiwaya2000tunnelling,kreisel2020remarkable,fernandes2022iron}. For clarity, only the fits for $G_1$ to $G_4$ (Solid lines) are displayed in Fig.~\ref{fig:directionalAR}c (the full set of fitting results and parameters are displayed in Extended Data Fig.~\ref{ExtenedFig:BTKfitARS}). Here, the value of the total gap ($\Delta_{total}$ = $\Delta_d$ +$\Delta_s$) determined by the fits is around 2.5 meV, and is nearly independent of Z, $\Gamma$ and $\alpha$ (Extended Data Fig.~\ref{ExtenedFig:BTKfitARS}d,e). The fitting further reveals $\Delta_d \gg \Delta_s$, suggesting the sOP is almost entirely $d$-wave. In contrast, the ARS spectra of bulk FeSe are quantitatively best described by an extended $s$-wave form (Extended Data Fig.~\ref{ExtenedFig:BTKfitARS}f). Interestingly, we note that the gap amplitude for the nodeless state ($\Delta_0$ +$\Delta_1$ in Extended Data Fig.~\ref{ExtenedFig:BTKfitARS}f) is roughly the same as $\Delta_{total}$ for the nodal state, suggesting similar pairing potentials. \rev{In addition, we considered whether the ZBCP could arise from an ABS of a pure $p$-wave superconducting state which also has a sign-changing nodal sOP.\cite{kashiwaya2000tunnelling,daghero2011directional} In particular, a $p$-wave gap, such as $p_x$ or $p_y$, has twofold angular symmetry, so two orthogonal crystal edges should not both correspond to equivalent nodal directions. However, pronounced ZBCPs without coherence-peak features are observed for contacts on two orthogonal Fe--Se-oriented edges (Extended Data Fig.~\ref{ExtenedFig:BTKfitARS}), indicating that the nodal/sign-changing structure repeats every $\pi/2$, disfavoring pure $p$-wave scenario.} 

Here, the reliability of the extracted $\alpha$ was confirmed by the strong agreement between the SEM results and the values extracted from the 2D-BTK fitting (as shown in Fig.~\ref{fig:directionalAR}c inset). Additionally, the experimentally determined angle-dependent, normalized, zero-bias conductances (see inset Fig.~\ref{fig:directionalAR}d) lie within reasonable bounds of theoretical curves for $0.4\leq Z \leq 0.6$, consistent with typical contact variations (see Extended Data Fig.~\ref{ExtenedFig:BTKfitARS}). Furthermore, the lifetime broadening of the ARS curves is exceptionally small compared to that in previous point-contact studies on FeSe-family flakes ($\Gamma \ll \Delta$, see Extended Data Fig.~\ref{ExtenedFig:BTKfitARS}d-f)\cite{ku2025point,naidyuk2017superconducting}. Meanwhile, once again, the temperature evolution of the $\alpha$-dependent anisotropic ARS is consistent with $T_{c}\approx 6 K$ measured from the four-point resistance, and $G_N$ is nearly temperature-independent (see Extended Data Fig.~\ref{ExtenedFig:TdependentARS}). These results once again confirm that the ARS response is truly spectroscopic and contacts are well within the ballistic regime.

\section*{Thickness Evolution}
\begin{figure}[t]
    \centering
    \includegraphics[width=0.5\linewidth]{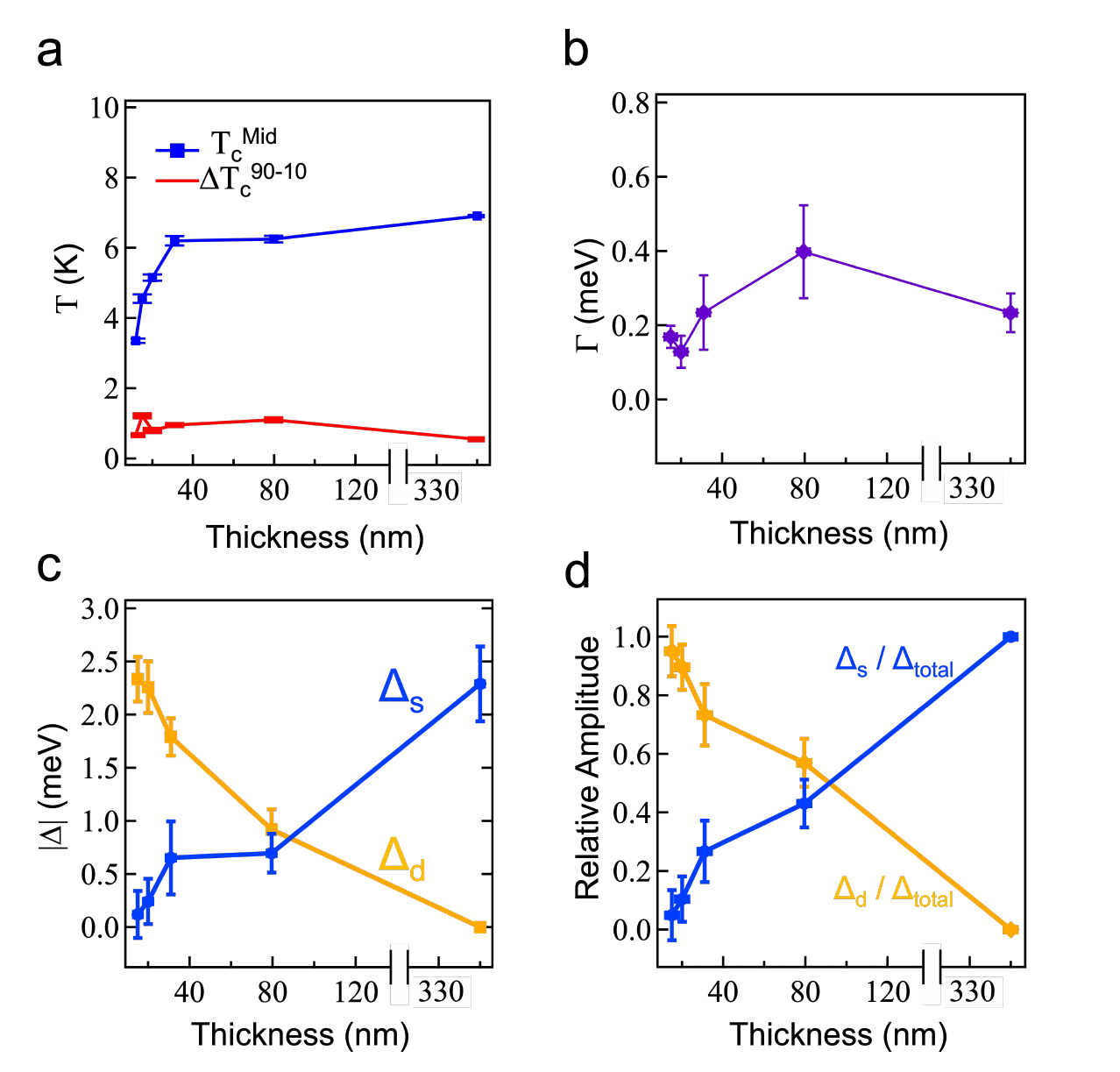}
    \caption{\label{Fig:ThicknessSummary}
    \textbf{Thickness evolution of superconducting transition, spectral broadening, and fitted gap components.} \textbf{a,} Thickness dependence of the superconducting transition. $T_c^{\mathrm{Mid}}$ is defined as the temperature where the resistance reaches $50\%$ of the normal-state resistance, and $\Delta T_c^{90-10}$ is defined as the temperature width between the $90\%$ and $10\%$ resistance criteria. To clarify, the $\Delta T_c^{90-10}$ is multiped 3 times. \textbf{b,} Dynes broadening parameter $\Gamma$ extracted from 2D-BTK fits to the ARS spectra versus thickness. \textbf{c,} Thickness dependence of the fitted superconducting gap amplitudes obtained from the 2D-BTK model. $\Delta_s$ and $\Delta_d$ denote the fitted $s$-wave and $d$-wave gap components, respectively.  \textbf{d,} Relative gap amplitude, defined as $\Delta_s/\Delta_{\mathrm{total}}$ and $\Delta_d/\Delta_{\mathrm{total}}$, where $\Delta_{\mathrm{total}}=\Delta_s+\Delta_d$. Here, for devices containing multiple ARS contacts, all 2D-BTK-derived quantities, including $\Gamma$, $\Delta_s$, and $\Delta_d$, are obtained by averaging over all measured contacts on the same device; the corresponding error bars are assigned conservatively as the largest fitting uncertainty among those contacts.
    }
\end{figure}

\rev{To further examine how the superconducting response evolves with reduced thickness, we summarized the available transport and 2D-BTK fitting results across devices with different flakes (Fig.~\ref{Fig:ThicknessSummary}). We first discuss the possible origin of the reduction of $T_c$ with decreasing thickness. In reduced-dimensional superconductors, enhanced phase fluctuations, stronger Coulomb repulsion, disorder scattering, and reduced phase stiffness can all suppress the global resistive $T_c$. Previous transport studies of exfoliated FeSe indeed reported that both $T_c$ and $T_s$ are suppressed with decreasing thickness, with a more pronounced superconducting suppression below 10 nm, and that Berezinskii--Kosterlitz--Thouless-like behavior appears in the few-layer limit.\cite{zhu2021evolution} Nevertheless, our ARS devices, with thicknesses from 15 to 350 nm, mostly lie above the few-layer regime where such two-dimensional fluctuation effects are expected to be most pronounced.}

\rev{As shown in Fig.~\ref{Fig:ThicknessSummary}a, $T_c^{\mathrm{Mid}}$ decreases with reduced thickness, while the superconducting transition width remains moderate and does not show a monotonic increase with reduced thickness or suppressed $T_c$. In addition, the Dynes broadening parameter $\Gamma$ extracted from the same 2D-BTK fits remains small compared with the superconducting gap scale and shows no systematic enhancement in thinner flakes or lower-$T_c$ devices (Fig.~\ref{Fig:ThicknessSummary}b). These trends are not consistent with the systematic enhancement expected if disorder or macroscopic inhomogeneity were the dominant origin of the $T_c$ suppression.\cite{daghero2011directional,kashiwaya2000tunnelling} Furthermore, we note that the decrease of $T_c$ can also be connected to the weakening of the bulk-like nematic response: as exfoliation suppresses the nematic reconstruction, FeSe is shifted away from the bulk-like pairing regime, which can reduce the effective pairing strength and global superconducting transition temperature.\cite{fernandes2013nematicity,kang2018superconductivity} Therefore, in our experiment the observed $T_c$ reduction likely reflects a combination of weakened bulk-like nematicity, modified pairing tendency, and generic reduced-dimensionality effects.}

\rev{We also note that the absence of a systematic enhancement of $\Gamma$, together with the moderate transition width, argues against severe macroscopic inhomogeneity or disorder-induced lifetime broadening as the primary origin of the modified ARS spectra and the ZBCP in thin flakes.\cite{ruf2024natural,dynes1978direct,daghero2011directional} Moreover, the reproducible crystallographic angle dependence, the disappearance of the ZBCP at $T_c$, and the magnetic-field response consistent with Doppler shift and vortex smearing further argue against impurity-induced in-gap states or disorder-induced lifetime broadening as the primary origin of the observed ZBCP (see Supplementary Information Section~XI).} 

\rev{Finally, we turn to study the superconducting order-parameter evolution. Because the absolute amplitude of the ZBCP can depend sensitively on contact barrier strength, normal-state background subtraction, spectral normalization, and the electronic effective temperature,\cite{kashiwaya2000tunnelling,daghero2011directional} we use the gap components extracted from the minimal 2D-BTK analysis (Extended Data Fig.~\ref{ExtenedFig:BTKfitARS}) as a more robust metric for evaluating the relative $s$- and $d$-wave pairing contributions. Following the minimal-model strategy (Supplementary Information Section II), the thin-flake spectra are fitted by the $s+d$ model, from which $\Delta_s$ and $\Delta_d$ are extracted, whereas the thick, bulk-like flake is well captured by an extended $s$-wave model without introducing a $d$-wave component. Notably, the 2D-BTK fits reveal a systematic thickness trend: thinner flakes are characterized by a large $d$-wave component and a smaller $s$-wave component (Fig.~\ref{Fig:ThicknessSummary}c).} 

\rev{This evolution is captured more directly by the relative gap amplitudes (Fig.~\ref{Fig:ThicknessSummary}d). For the bulk-like extended-$s$ fit, the $d$-wave component is not introduced in the minimal model and is therefore taken as zero in this relative-amplitude comparison. Within the available device set, the relative $d$-wave weight increases strongly upon reducing thickness, while the $s$-wave weight becomes dominant in the thick, bulk-like limit. These results support a thickness-driven evolution from a bulk-like $s$-wave-dominant state toward a $d$-wave-dominant state in exfoliated FeSe, while leaving the detailed crossover form and critical thickness scale for future systematic studies.}

\section*{Discussion}
Our combined Raman and transport measurements indicate that reduced dimensionality strongly weakens the bulk-like nematics in FeSe\cite{farrar2020suppression,zhu2021evolution,xie2021fragile}, which surprisingly enables the emergence of a $d$-wave-dominant sOP as probed by ARS. For comparison, we note that exfoliation in an analogous material, FeTe${0.55}$Se${0.45}$, was reported to induce a dramatic $c$-axis expansion ($\sim$18\%) and to alter the lattice symmetry.\cite{kong2025cooper} To examine whether exfoliated FeSe undergoes similar structural modifications that might also underlie its nematic and superconducting transformation, we performed scanning transmission electron microscopy (STEM; Extended Data Fig.~\ref{ExtenedFig:STEM-HAADF}a-e). These measurements (Extended Data Fig.~\ref{ExtenedFig:STEM-HAADF}c) reveal that the average c-axis parameter of thin flakes is not distinct from that of bulk (5.5 Å). Next, helicity- and linear-polarization-resolved Raman measurements were carried out. The angular dependence of the Raman response in exfoliated flakes closely matches that of bulk crystals (Supplementary Fig.~S5), indicating no global change in lattice symmetry (Supplementary Section~VII). 

\rev{However, although STEM and polarization-resolved Raman rule out a dramatic structural or lattice-symmetry deformation, more subtle lattice modifications after exfoliation remain possible. These may include small changes in the in-plane lattice constants, local structural disorder, changes in the Se height, or modifications of the Se-Fe-Se bond angle. These possibilities are motivated by prior work showing that the electronic structure and nematicity of FeSe are highly sensitive to strain and lattice tuning,\cite{phan2017effects,leonov2015correlation,skornyakov2018correlation} and that exfoliated or substrate-supported two-dimensional materials can retain local strain or relax it through wrinkles and local lattice reconstruction.\cite{li2021lattice,halbertal2023multilayered,basu2023strain} We therefore performed DFT calculations to assess how the electronic structure of FeSe evolves under potential lattice relaxation (Extended Data Fig.~\ref{ExtenedFig:DFT}). In these calculations, the unrelaxed tetragonal FeSe band structure is compared with structures in which the lattice constants are fixed while the internal atomic positions are relaxed using the custom relaxation procedure (Supplementary Information Section~IX). The full set of calculations spans lattice parameters $a=b=3.79$--$4.00~\angstrom$ and $c=5.525$--$6.05~\angstrom$, as shown in Supplementary Figs.~S6-8.}

\rev{These results suggest a plausible microscopic route to weaken the bulk nematic instability. First, the inner $\Gamma/Z$-centered band shifts systematically downward as the lattice expands, with a corresponding evolution near $Z$. This downward shift can reduce the phase space for low-energy hole-electron pocket scattering between $\Gamma/Z$ and $M$, thereby weakening the orbital-fluctuation channels that help stabilize the bulk nematic state.\cite{massat2016charge,gallais2016nematic,steffensen2021interorbital} Because FeSe has very small Fermi pockets, a modest band shift can drive a Lifshitz-like change of the hole pocket (Extended Data Fig.~\ref{ExtenedFig:DFT}d) around $a=3.94$~\angstrom~(Supplementary Fig.~S8). This strongly modifies the Fermi-surface topology and reduces the low-energy orbital spectral weight contributing to the nematic susceptibility, without requiring a large change in overall carrier density. A related precedent is found in FeSe$_{1-x}$S$_x$, where quantum-oscillation measurements found that a small, highly mobile orbit disappears near the nematic phase boundary, indicative of a Lifshitz transition.\cite{coldea2019evolution} Finally, the orbital-resolved DFT calculations (Supplementary Information Section~IX) indicate that the $\Gamma/Z$-centered inner band that shifts downward has mixed Fe $d_{x^2-y^2}$ and $d_{xz}/d_{yz}$ orbital character\cite{zhang2015observation,yi2019nematic}. Since, the bulk FeSe nematicity is associated with lifting of the $d_{xz}/d_{yz}$ degeneracy and a strong reconstruction of the low-energy Fermi surface, pushing this mixed-orbital band away from $E_F$ therefore reduces the
low-energy orbital spectral weight that may contribute to bulk
nematicity.\cite{shimojima2014lifting,watson2015emergence,yi2019nematic}}

\rev{Next, we ask how such a change in nematicity can alter the superconducting pairing symmetry. In Fe-SC in general and FeSe in particular, the $s$-wave and $d$-wave pairing channels are known to be closely competing, and their relative stability is highly sensitive to the nematic reconstruction of the Fermi surface and orbital content.\cite{graser2009near} To examine this connection, we performed a complementary theoretical analysis based on a full five-orbital, two-sublattice FeSe model (Supplementary Information Section~X). To study the impact of nematicity on the superconducting pairing, interactions were introduced in the nearest-neighbor $B_{1g}$ nematic channel of the unfolded Brillouin zone, and nematic mean-field order parameters $\Delta^\mathrm{nem}$, were obtained self-consistently as a function of the nematic interaction strength $g_{\mathrm{nem}}$ (Extended Data Fig.~\ref{ExtenedFig:RPA}a and Supplementary Information Section~X). For each self-consistently determined nematic Fermi surface, we then solved the linearized superconducting gap equation within the spin/charge-fluctuation RPA framework (Supplementary Information Section~X).\cite{graser2009near,saito2015revisiting,Rhodes2021nonloc}}

\rev{This calculation shows that nematicity acts as a tuning knob between the competing superconducting pairing channels. In the non-nematic regime, $g_{\mathrm{nem}}<390$~meV, the leading superconducting instability is $d$-like (Extended Data Fig.~\ref{ExtenedFig:RPA}b). Once a finite nematic mean-field order develops, the breaking of four-fold rotational symmetry allows $s$- and $d$-like components to mix. As $g_{\mathrm{nem}}$ increases, the nematic order grows, the Fermi surface becomes increasingly distorted (see Supplementary Fig.~S9), and the projected $s$-like weight of the leading gap solution increases (see Supplementary Fig.~S10). In the weakly nematic regime, $390 <g_{\mathrm{nem}}< 401$~meV, the state remains largely $d$-like but acquires an increasing $s$-wave admixture. At stronger nematic coupling, the $s$-like component becomes dominant, consistent with the nodeless bulk-like spectra. Thus, this model supports the trends that we observed in the ARS spectra in exfoliated FeSe flakes.}

\rev{Therefore, the DFT and RPA calculations provide a coherent framework linking exfoliation, nematicity, and superconducting pairing symmetry. The DFT results suggest that subtle lattice relaxation can selectively reconstruct the $\Gamma/Z$-centered low-energy bands and drive a Lifshitz-like change of the hole pocket, thereby weakening the Fermi-surface and orbital conditions that stabilize the bulk nematic state. The RPA calculation then shows that weakening nematicity shifts the balance between competing pairing channels from a bulk-like $s$-wave-dominant state toward a $d$-wave-dominant state. This picture naturally connects the Raman, transport, and ARS spectra in exfoliated FeSe.} 

In summary, our combined electrical transport, Raman, and angle-dependent ARS on crystal edges confirmed that reducing the thickness of FeSe by exfoliation produces dramatic changes in its electronic and superconducting orders. Without chemical substitution or external strain, we realize a transition from a nodeless to a sign-changing nodal superconducting state, accompanied by a strong weakening of nematicity and the absence of a bulk-like nematic-transition signature. Thus, exfoliation of FeSe bridges the phenomenology of nodeless and nodal superconductors and could enable precise engineering of the delicate interplay between nematicity, orbital degrees of freedom, and reduced dimensionality. Whereas chemical substitution, which often leads to increased inhomogeneity\cite{cho2019strongly}, and high-pressure setups limit the application of local probe methods such as STM\cite{pei2022high}, our technique enables a fresh approach to quantifying various factors in the long-debated superconducting pairing in FeSCs\cite{kreisel2020remarkable,fernandes2022iron}. This further supports that the exfoliated FeSe family provides a versatile platform for designing and discovering new superconducting phenomena\cite{ishida2022pure,kong2025cooper}. Indeed, FeSe presents a new opportunity to explore nematicity and $d$-wave order without the complications of exfoliated Bi$_2$Sr$_2$CaCu$_2$O$_8$\cite{zhao2023time,zareapour2017andreev, yu2019high}, and with the possibility to create new topological states in M-point Moire structures\cite{garcia2025symmetry}. Furthermore, our demonstration of reliable ARS in 2D materials combined with Raman could uncover new superconducting transitions or intertwined quantum states in other van der Waals superconductors\cite{Xi.2015,kvashnin2020coexistence}, Fe-based derivatives\cite{kreisel2020remarkable,fernandes2022iron}, and even in nickelates\cite{wang2020distinct}, cuprates\cite{auvray2019nematic} or kagome systems\cite{song2023anomalous} where nematic, CDW or orbital instabilities coexist with superconductivity. Lastly, exfoliated FeSe opens avenues for designing symmetry-tunable quantum devices and heterostructures that exploit unconventional or topological superconducting states. 

\bibliographystyle{naturemag}
\bibliography{Reference.bib}

\section*{Data and materials availability} 
All data are available in the manuscript or the Supplementary Information.
\section*{Acknowledgments}
We are grateful to R. Rafael Fernandes and D. Morr for helpful discussions. We thank Kyle Fruhling and Fazel Tafti for assistance with the magnetic susceptibility measurements.
\noindent\textbf{Funding:} The data taking, analysis, manuscript preparation, and fabrication efforts of W.L. and K.S.B. are based upon work supported by the Air Force Office of Scientific Research under award number FA2386-24-1-4071. Device fabrication efforts by G.N. and M.G. were supported by the National Science Foundation, Award No. DMR-2310895. The Raman experiments and analysis by B.S. and K.-M.K. were supported by the U.S. Department of Energy (DOE), Office of Science, Basic Energy Sciences (BES) under Award \# SC0018675. The research of M.P. was funded in part by The Robert A. Welch Foundation, Grant \# L-E-0001-19921203. Structure characterizations using electron microscopy techniques (S.D.F, J.J.C.) were supported by the DOE BES under Award \#DE-SC0023905. Work in Japan was supported by a Grant-in-Aid for Transformative Research Areas (A) ``Correlation Design Science'' (No. JP25H01248), and by Grants-in-Aid for Scientific Research (KAKENHI) (Nos. JP22H00105 and JP24K17007) from Japan Society for the Promotion of Science. We acknowledge the support of the Natural Sciences and Engineering Research Council of Canada (NSERC) and the support of grant 369963 from the Fonds de recherche du Qu\'ebec for the numerical works in the article.

\noindent\textbf{Author contributions}  
K.S.B. conceived and supervised the project; FeSe devices are prepared by W.L. with the support of G.N. and M.G.; W.L. designed and conducted the transport and ARS experiments, with the help of G.N.; K.-M.K., B.S.  performed the Raman measurements and analyzed the data, and P.S. assisted with data collection; S.D.F. and J.J.C. performed the STEM and EBSD measurement; The DFT results are calculated by A.D., A.D.L., and M.G.V.; K.I. performed the fit of Hall data. W.L., K.S.B., and M.P. were responsible for the main discussion of the results. FeSe materials are provided by L.K. and T.S.; The manuscript is mainly drafted by W.L. with the help of K.S.B., M.P., G.N., K.-M.K., A.D., A.D.L., B.S., K.I.; All authors contributed to the discussion of the manuscript. 

\section*{Competing interests} 
The authors declare no competing interest. 

\section*{Methods}
\subsection{Exfoliation and Device Fabrication}
Although FeSe single crystal is stable in air, exfoliated thin flakes are very air-sensitive. Careful pre-fabrication characterization of mechanically exfoliated FeSe flakes was required to properly fabricate devices to enable the detection of the optical and transport spectra. All device fabrication and characterization from exfoliation to metal deposition was performed in our unique 'Clean-room in a Glovebox,' which allows us to make complex devices without exposing samples to ambient conditions\cite{gray2020cleanroom}. The Single-crystalline FeSe was synthesized by chemical vapor transport methods in the University of Tokyo. Bulk FeSe crystals were mechanically exfoliated onto \ch{Si/SiO_{2}} and sapphire substrates. The edges of the FeSe single crystal were naturally parallel to [100]/[010] in the most cases, and randomly along crystalline axes with other directions, which promise the presence of ultra straight edges. Using optical microscopy, the FeSe flakes were carefully searched for the feature of two sharp orthogonal edges. Here, via SEM-EBSD, we confirmed two orthogonal sharp edges on FeSe flakes are along <100> direction as expected (see Supplementary Fig.~S2). 

Before fabrication, candidate flakes are further evaluated using AFM to assess the sharpness of the edges, ensure they are at orthogonal angles, flake thickness, and confirm the absence of defects or warping of the flake surface. Direct-write photolithography was then employed to pattern contacts (width $\sim$ 0.5 $\mu$m and length $\sim$3 $\mu$m) before the sample was loaded into an thermal evaporation or electron-beam evaporator with in situ plasma and pumped down to pressure $\sim1\times10^{-7}$ torr. Most devices employed used a 5 nm Cr wetting layer followed by 45 nm of Au, but for thick flakes the thickness of Au deposition is correspondingly enlarged. We didn't find much difference of metal deposition between thermal evaporation and e-beam here. Here, contacts for Andreev spectroscopic measurement were fabricated as twin-branch design, in which they split out into separate current (I+) and voltage (V+) leads as close to the flake edge as possible. This limits the additional resistance contributions from the metallic lead itself: namely, the resistance of leads is around 0.1 $\Omega$ as calculated from the geometry and resistivity of leads, however, the normal-state resistance of ARS is in the range of 5 $\Omega$ to 46 $\Omega$ (see Supplementary Fig.~S3), meaning most of the resistance is contributed by the interfacial contacts.

\rev{Even inside the glovebox, the surface of thin FeSe flakes can gradually accumulate residual contamination or weak oxidation during processing. We therefore applied an optimized, brief in-situ Ar plasma treatment immediately before metal deposition to remove residual resist and surface/edge contamination and to improve the normal-metal--FeSe interface quality. Ar plasma is different from O$_2$ plasma because it does not chemically oxidize FeSe; instead, it primarily cleans the exposed surface through physical ion-assisted removal. However, excessive Ar plasma exposure can also damage thin flakes through ion bombardment, physical sputtering, near-surface disorder, or possible chalcogen depletion.\cite{zhu2017argon} Therefore, the plasma condition was carefully optimized rather than treated as a routine cleaning step. This optimization was particularly important for edge-contact ARS measurement. Additionally, we found that an appropriate Ar plasma treatment more readily exposes the sidewall contact path than the top-surface path, enabling the current injection to be dominated by the FeSe side surface rather than by the oxidized or contaminated top surface. Underexposure, either from too low a plasma power or too short an exposure time, typically resulted in poor lift-off, noisy signals, or M$\Omega$-scale contact resistance, consistent with incomplete removal of sidewall contamination or oxide. In contrast, overexposure produced devices with degraded contact behavior in which contacts did not show low-resistance ballistic Sharvin behavior (i.e., small effective contact sizes and low Dynes broadening), and therefore no reproducible spectroscopic Andreev features were observed. We optimized the process by fabricating and characterizing a series of devices with different Ar plasma powers and exposure times. A 60~s, 60~W Ar plasma treatment was selected because it reproducibly produced Ohmic, spectroscopic contacts with normal-state resistances in the ARS range and clear superconducting low-bias features that disappear at $T_c$ measured by four-point contacts that probe the bulk.}

\subsection{Symmetry-resolved Raman spectroscopy}

\rev{To minimize degradation of the air-sensitive exfoliated FeSe samples, the crystals were freshly cleaved and exfoliated inside a glovebox and transferred directly under high vacuum ($1.0 \times 10^{-8}$ Torr) to the low-temperature Raman cryostat using a vacuum suitcase.\cite{gray2020cleanroom} Raman measurements were performed in a back-scattering geometry using a 532-nm (2.33-eV) excitation laser focused through a 100$\times$ objective. The scattered light was dispersed by a 1200-grooves/mm grating and collected with an Andor spectrometer. The incident laser power was kept below 0.2 mW to minimize local heating. The residual laser-heating effect was independently evaluated using the Stokes-to-anti-Stokes intensity ratio, as discussed in Supplementary Information Section~VIII.\cite{wang2020range} The different Raman polarization configurations were selected by controlling the incident and scattered-light polarizations using half-wave plates and a fixed analyzer; circular-polarization measurements employed quarter-wave plates. Temperature-dependent measurements were performed under high vacuum ($\sim1.0\times10^{-8}$ Torr) using a closed-cycle cryostat.}

\rev{The measured Stokes spectra were corrected for the Bose thermal factor to obtain the Raman response $\chi''_{\mu}(\omega,T)$, where $\mu$ denotes the polarization configuration. To quantify its temperature evolution, we evaluate a finite-window partial dynamic susceptibility using the Kramers--Kronig relation,
\[
\chi_{\mu}^{\mathrm{part}}(T;\omega_{\min},\Omega)
=
\frac{2}{\pi}
\int_{\omega_{\min}}^{\Omega}
\frac{\chi''_{\mu}(\omega,T)}{\omega}\,d\omega ,
\]
where $\omega_{\min}$ and $\Omega$ denote the experimentally accessible lower bound and the selected upper integration cutoff, respectively. Because the measured spectra do not extend continuously to zero frequency or infinite energy, $\chi_{\mu}^{\mathrm{part}}$ represents a finite-window partial susceptibility and is not identified with the complete zero-frequency dynamic susceptibility.}

\rev{For the bulk measurements, the accessible low-energy limit of the measurement configuration is approximately 5 meV. We selected the upper cutoffs to progressively extend the integration from the low-energy regime dominated by the critical QEP to the broader higher-energy $B_{1g}$ response established in previous Raman measurements of FeSe.\cite{massat2016charge,zhang2021quadrupolar,chibani2021lattice} The susceptibility shown in the main text is evaluated up to 50 meV, which captures the principal low-energy nematic spectral evolution and spans the characteristic energy scale of the nematic electronic reconstruction in FeSe.\cite{yi2019nematic} To test the dependence on the integration range, additional upper cutoffs of 25, 80, and 150 meV were evaluated for the bulk $B_{1g}$ response (Extended Data Fig.~\ref{ExtenedFig:BulkRaman}e). The pronounced maximum near $T_s$ persists across all tested cutoffs, showing that the identified transition feature is robust to substantial changes in the upper integration limit. The corresponding $B_{2g}$ susceptibilities evaluated with upper cutoffs of 25, 50, and 150 meV remain smooth and show no analogous maximum (Extended Data Fig.~\ref{ExtenedFig:BulkRaman}f).}

\rev{For the exfoliated-flake measurements, the experimentally accessible low-energy limit is approximately 2 meV. The main thin-flake partial dynamic susceptibilities are evaluated up to 50 meV (Extended Data Fig.~\ref{ExtenedFig:FlakeRaman}e,f). The 50-meV upper bound captures the low-energy response while remaining below the dominant first-order Si Raman line near 65 meV. We tested the robustness of the extracted temperature dependence using several alternative low-energy integration windows. In addition, a masked wide-energy result combines the 2--57 and 74--123-meV intervals, omitting the range containing the dominant Si Raman line (Supplementary Fig.~S11). We note that this masked result is used only as an integration-window sensitivity test. No empirical substrate subtraction or temperature-dependent intensity rescaling was applied.}

\subsection{Differential Conductance Measurements}
The differential conductance measurements were performed at a base temperature of $\sim$ 1.4 K in a Liquid Helium Sub-cooled Variable Temperature Insert with 9 T Superconducting Magnet (Cryo Industries of America Inc.) at Boston College with out-of-plane sample rotation. Generally, a current-biased AC modulation technique was used to measure the differential conductance ($G=dI/dV$) as a function of the measured bias voltage at the interface between the FeSe sample and the normal-metal lead. A DC current (Keithley 6221) is modulated by a small AC current generated by using a lock-in amplifier (Stanford Research Systems SR830) to output an AC voltage (typically $\sim$ 0.5 - 0.8 V, 573.73 Hz or 1777.77 Hz) across a 1 MOhm resistor. A custom box is then used to join the AC and DC components before passing them through the contact of interest. The resulting DC voltage across the point contact is measured using a digital multimeter (Agilent AG34401a), and the AC voltage is first passed through a voltage pre-amplifier (Stanford Research Systems SR560), which is used only for its filtering capabilities, and then measured using the SR830 lock-in amplifier. The dI/dV is calculated from the lock-in's AC current amplitude (dI) and the measured AC voltage (dV). All data collection was performed using a custom LabVIEW interface. 

\subsection{Hall effect Measurements}
Bulk FeSe is a multi-band materials with hole-like Fermi surfaces at $\Gamma$ point, and two electron-like Fermi surfaces at $M$ point. Correspondingly, the Hall response will reflect its multi-carrier characteristics. The Hall resistivity $\rho_{xy}(B)$ and magnetoresistivity $\rho_{xx}(B)$ are probed by four-terminal measurement configuration (see Supplementary  Fig.~S4a) at $T=15\,\mathrm{K}$, referenced by the previous magnetotransport measurement on FeSe\cite{watson2015dichotomy}. Generally, the changing of feature in  $\rho_{xy}$ (i.e., B-nonlinear to B-linear) implies a distinct Fermiology and/or dramatic change in the relative mobilities or densities of the carriers.  To explore this, a 2-band and a 3-band models were used to extract carrier densities and mobilities.

Here, the linear, positive $\rho_{xy}(B)$ in thin flakes suggest a compensated 2-band system\cite{watson2015dichotomy}. The simplest 2-bands model of this system can be described as:
$
\rho_{xy} = \frac{B}{e} \cdot 
\frac{(n_h \mu_h^2 - n_e \mu_e^2) + (n_h - n_e)\mu_h^2 \mu_e^2 B^2}
{(n_h \mu_h + n_e \mu_e)^2 + (n_h - n_e)^2 \mu_h^2 \mu_e^2 B^2}
$, where $n_{e,h}$ is the carrier density and $\mu_{e,h}$ is the carrier mobility. The $\rho_{xx}(B)$ is determined by balancing the carrier densities and mobilities between holes and electrons. In bulk FeSe where the nematic reconstruction ($T < T_s$) yields high-mobility Dirac-like electrons at the $M-$point coexisting with holes at $\Gamma$, producing nonlinear and negative $\rho_{xy}$, the description goes beyond the 2-bands compensated model and requires a 3-band model with one hole and two electron bands.\cite{watson2015dichotomy} The conductivity tensors in this model can be expressed as:
$
\sigma_{xx} = \frac{\rho_{xx}}{\rho_{xx}^2+\rho_{xy}^2} = \frac{n_h e \mu_h}{1+\mu_h^2B^2} + \frac{n_{e1} e \mu_{e1}}{1+\mu_{e1}^2B^2} + \frac{n_{e2} e \mu_{e2}}{1+\mu_{e2}^2B^2}
$
$
\sigma_{yx} = \frac{\rho_{xy}}{\rho_{xx}^2+\rho_{xy}^2} = \frac{n_h e \mu_h^2 B}{1+\mu_h^2B^2} - \frac{n_{e1} e \mu_{e1}^2 B}{1+\mu_{e1}^2B^2} - \frac{n_{e2} e \mu_{e2}^2 B}{1+\mu_{e2}^2B^2}
$, where $n_i$ and $\mu_i$ represent the carrier density and mobility in each band $i$, respectively, and $e$ is the elementary charge. By assuming that the electric transport is isotropic, in the 2-band fit, we put $n_{e2} = 0$ and optimized $n_i$ and $\mu_i$ ($i$ = h and e1). In the 3-band model fit, we simultaneously fit $\sigma_{xx}(B)$ and $\sigma_{yx}(B)$ by optimizing $n_i$ and $\mu_i$ ($i$ = h, e1, and e2) with the carrier compensation condition, $n_h = n_{e1} + n_{e2}$. 

\subsection{Electron microscopy} 
Scanning transmission electron microscopy (STEM) was employed to characterize the lattice orientation and parameters of FeSe flakes. Electron transparent cross-sectional samples were prepared with a Thermo-Fisher Helios G4-UX Dual-Beam (SEM/FIB) with final thinning at 2 kV. A ThermoFisher Spectra aberration-corrected STEM microscope operating at 300 keV was used for atomic resolution high angle annular dark-field (HAADF) imaging, with a probe convergence semi-angle of 21.4 mrad and HAADF collection angle of 60-200 mrad. Images were acquired with dwell times of 2 $\mu$s per pixel and a beam current of 40 pA. For quantitative structural analysis, HAADF STEM image pairs were acquired at orthogonal scan directions. The image pairs were drift corrected using a non-linear drift correction code. [10.1016/j.ultramic.2015.12.002] Reference images of [110] silicon were taken to measure and correct instrumental scan distortion. To allow accurate comparisons with other structural probes, the image pixel size was calibrated from the fully (drift and scan distortion) corrected silicon image using the standard lattice parameter of silicon, 5.431020511(89)  [https://physics.nist.gov/cgi-bin/cuu/Value?asil]. Atomic scale structural measurements of FeSe flakes were performed on the fully corrected images using the Python package SingleOrigin [https://github.com/sdfunni/SingleOrigin]. This software performs 2D gaussian fitting to measure the positions of atomic columns in STEM images. SEM \& EBSD data were taken on a Zeiss Sigma 500 operated at 20 kV. The highest quality EBSD patterns were selected from scans of each flake and indexed manually.

\subsection{Magnetic Properties Measurement System} %
Magnetization measurements were performed on single crystals using an 8 T Quantum Design MPMS-3 equipped with vibrating sample magnetometry (VSM) and oven options. The MCE was quantified by measuring magnetization isotherms in the vicinity of the critical temperature and evaluating the magnetic entropy change using the Maxwell relation.

\section*{Supplementary Information}

Supplementary Information is available for this paper and includes additional experimental details, theoretical calculations, and Supplementary Figs. S1--S11.
\newpage
\section*{Figure Legends}

\newpage


\newpage

\newpage

\newpage

\newpage

\newpage
\begin{extendedfigure}[t]
    \centering
    \includegraphics[width=0.70\linewidth]{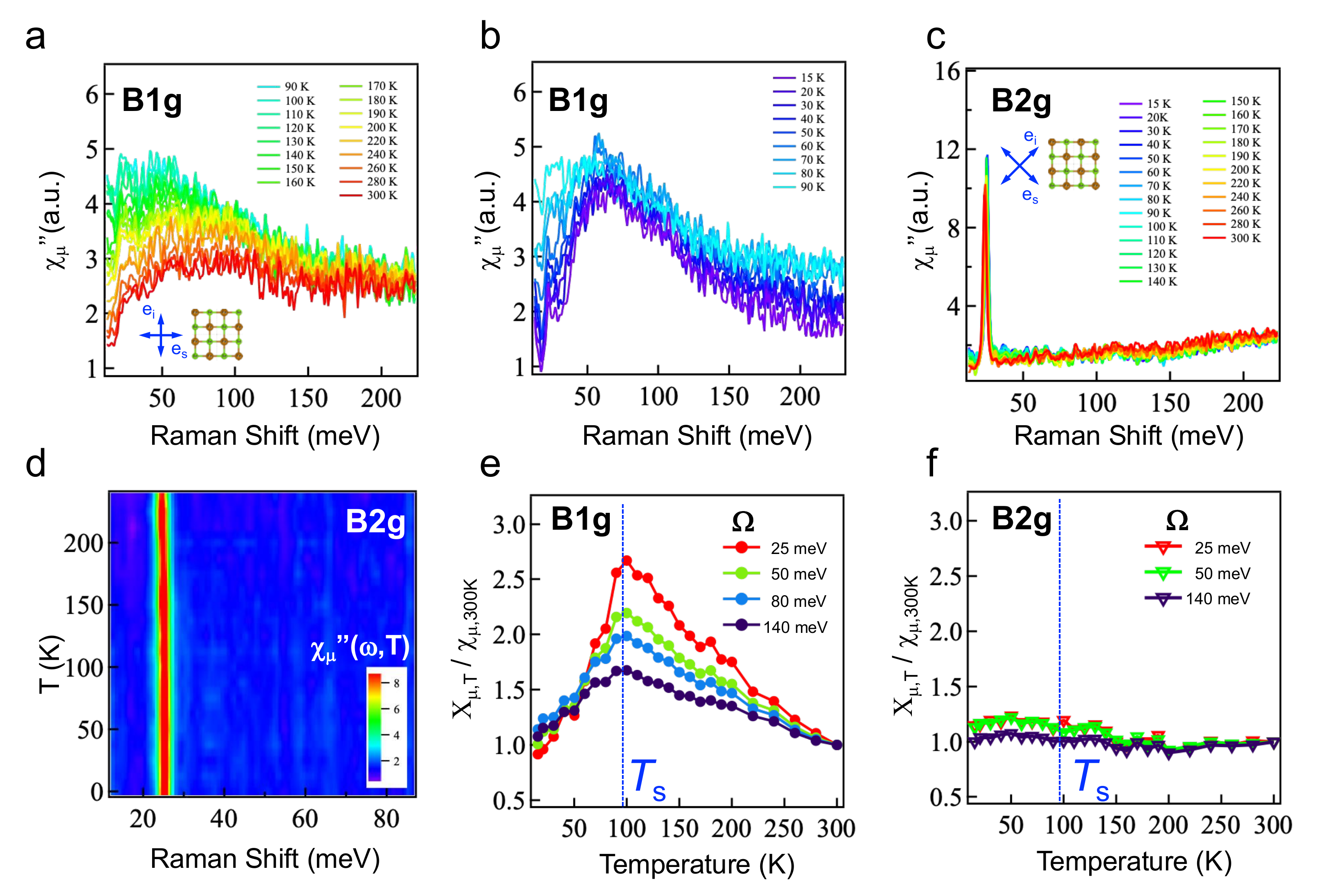}
    \caption{\label{ExtenedFig:BulkRaman} \textbf{Symmetry-dependent Raman spectra in bulk FeSe.} \textbf{a,} Temperature evolution of the bulk $B_{1g}$ Raman response, $\chi''_{B_{1g}}(\omega,T)$, from 90 to 300 K. \textbf{b,} Corresponding $B_{1g}$ Raman response from 15 to 90 K. \textbf{c,} Temperature evolution of the bulk $B_{2g}$ Raman response from 15 to 300 K. Inset: Schematic of the $B_{1g}$ symmetry. $B_{1g}$ is selected via the crossed incoming ($e_{i}$) and scattered ($e_{s}$) photon polarizations at 45 degrees relative to the Fe–Fe bonds. \textbf{d,} Temperature--Raman-shift map of the bulk $B_{2g}$ response. In contrast to the $B_{1g}$ response in Fig.~\ref{fig:Nematicity}b, no pronounced transition-like redistribution develops near $T_s$. \textbf{e,} Normalized finite-window partial $B_{1g}$ dynamic susceptibility calculated with a fixed lower integration limit of 5 meV and upper cutoffs $\Omega=25$, 50, 80, and 150 meV. Each curve is normalized to its value at 300 K. \textbf{f,} Corresponding normalized $B_{2g}$ partial dynamic susceptibilities for $\Omega=25$, 50, and 150 meV. No corresponding maximum near $T_s$ develops for any tested cutoff. Vertical dashed lines in \textbf{e} and \textbf{f} mark the bulk $T_s$.
    }
\end{extendedfigure}

\newpage
\begin{extendedfigure}[t]
    \centering
    \includegraphics[width=0.90\linewidth]{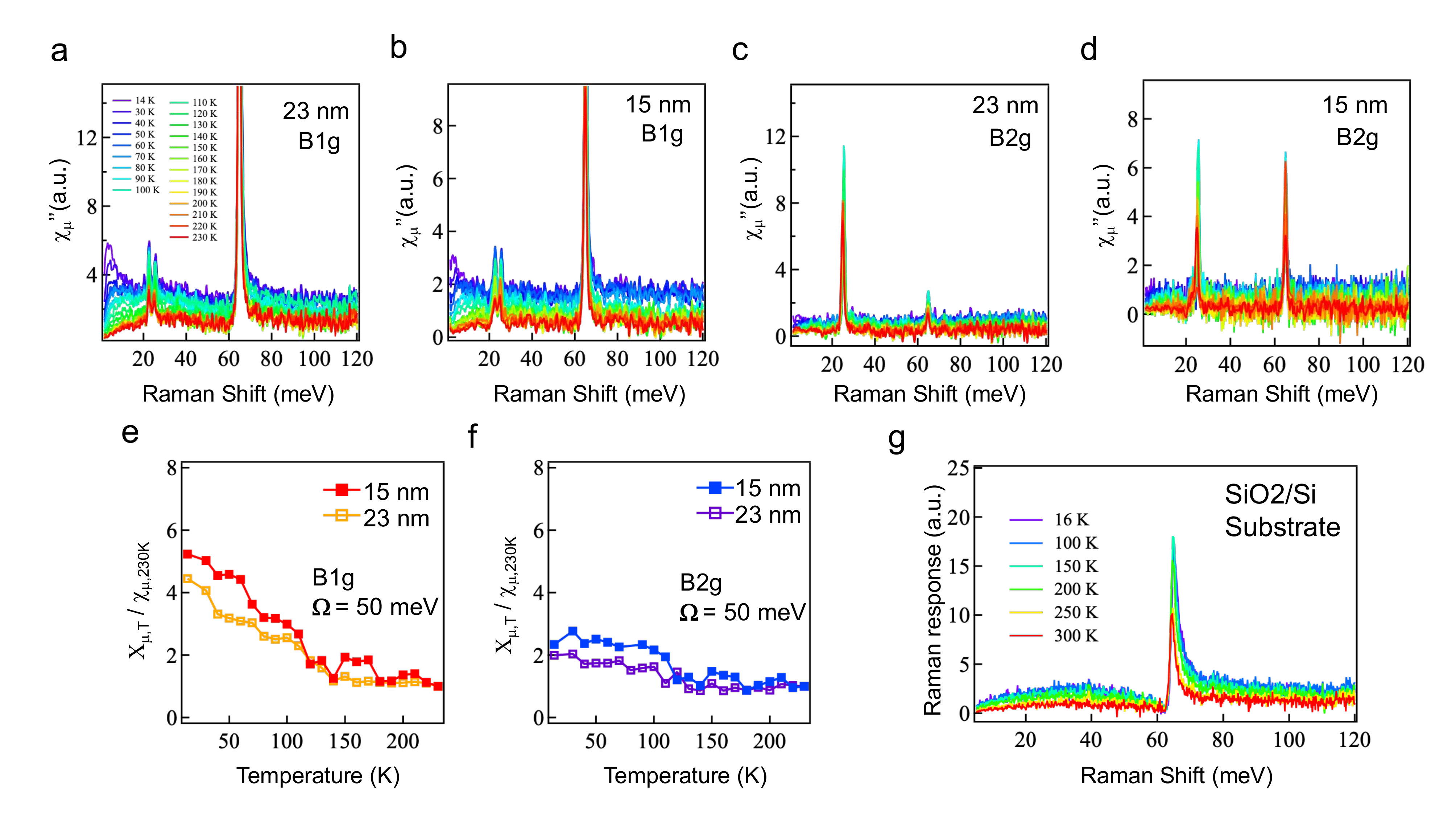}
    \caption{\label{ExtenedFig:FlakeRaman} \textbf{Temperature-dependent Raman response in exfoliated FeSe flakes.} \textbf{a-b,} Temperature evolution of the $B_{1g}$ Raman response, $\chi''_{B_{1g}}(\omega,T)$, for \textbf{a,} 23-nm and \textbf{b,} 15-nm FeSe flake. \textbf{c-d,} Corresponding $B_{2g}$ Raman response for \textbf{c,} 23-nm and \textbf{d,} 15-nm flakes. \textbf{e,} Normalized dynamic susceptibility in the $B_{1g}$ configuration for both flakes, normalized to the corresponding value at 230 K. \textbf{f,} Corresponding susceptibilities in the $B_{2g}$ configuration. \textbf{g,} Representative temperature-dependent Raman response of an independently measured SiO$_2$/Si substrate.    
}
\end{extendedfigure}

\newpage
\begin{extendedfigure}[t]
    \centering
    \includegraphics[width=0.8\linewidth]{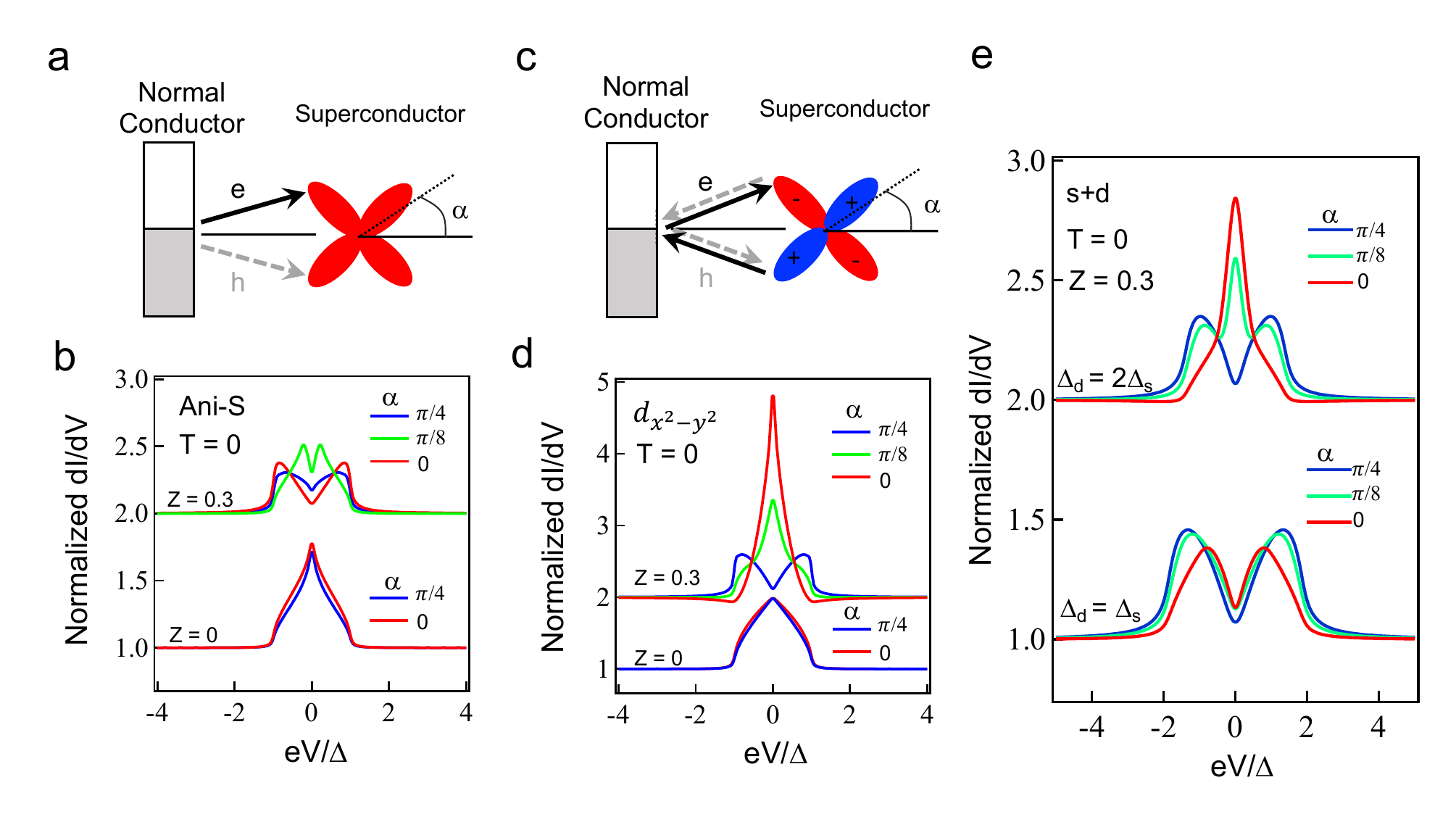}
    \caption{\label{ExtenedFig:SymmetryARSModel} \textbf{Symmetry-dependent Andreev spectra of the order parameter.} \textbf{a,} Schematic of the Andreev process in an extremely anisotropic but uniform phase superconductor (e.g., extended $s$-wave) \textbf{b,} 2D-BTK conductance curves for ARS on the extended $s$-wave superconductor, with Z = 0 and 0.3, at T = 0 K. The extended $s$-wave gap function is $\Delta_{es}(\mathbf{\alpha}) = \Delta_0 + \Delta_1 \cos(4\alpha+\pi)$, where $\Delta_0 = \Delta_1$ gives nodal but still sign-preserving sOP. The curves with Z = 0.3 are offset to 2 for clarity. \textbf{c,} Andreev process in $d$-wave superconductor which has anisotropic and sign-changing order parameter. \textbf{d,} Corresponding ARS of 2D-BTK model for a $d$-wave superconductor with interfacial barrier Z = 0 and 0.3. The gap function is $\Delta_{d}(\mathbf{\alpha}) = \Delta_d \cos(2\alpha+\frac{\pi}{2})$. The curves with Z = 0.3 are offset again to 2 for clarity. \textbf{e,} Normalized conductance curves at T = 0 K calculated with the 2D-BTK model for a $s$+$d$-wave superconductor $\Delta_{s+d}(\mathbf{\alpha}) = \Delta_s + \Delta_d \cos(2\alpha+\frac{\pi}{2}),$ with Z = 0.3, for sign-changing ($\Delta_s < \Delta_d$) or sign-preserving ($\Delta_s \ge \Delta_d$) case.}
\end{extendedfigure}

\newpage
\begin{extendedfigure}[t]
    \centering
    \includegraphics[width=0.9\linewidth]{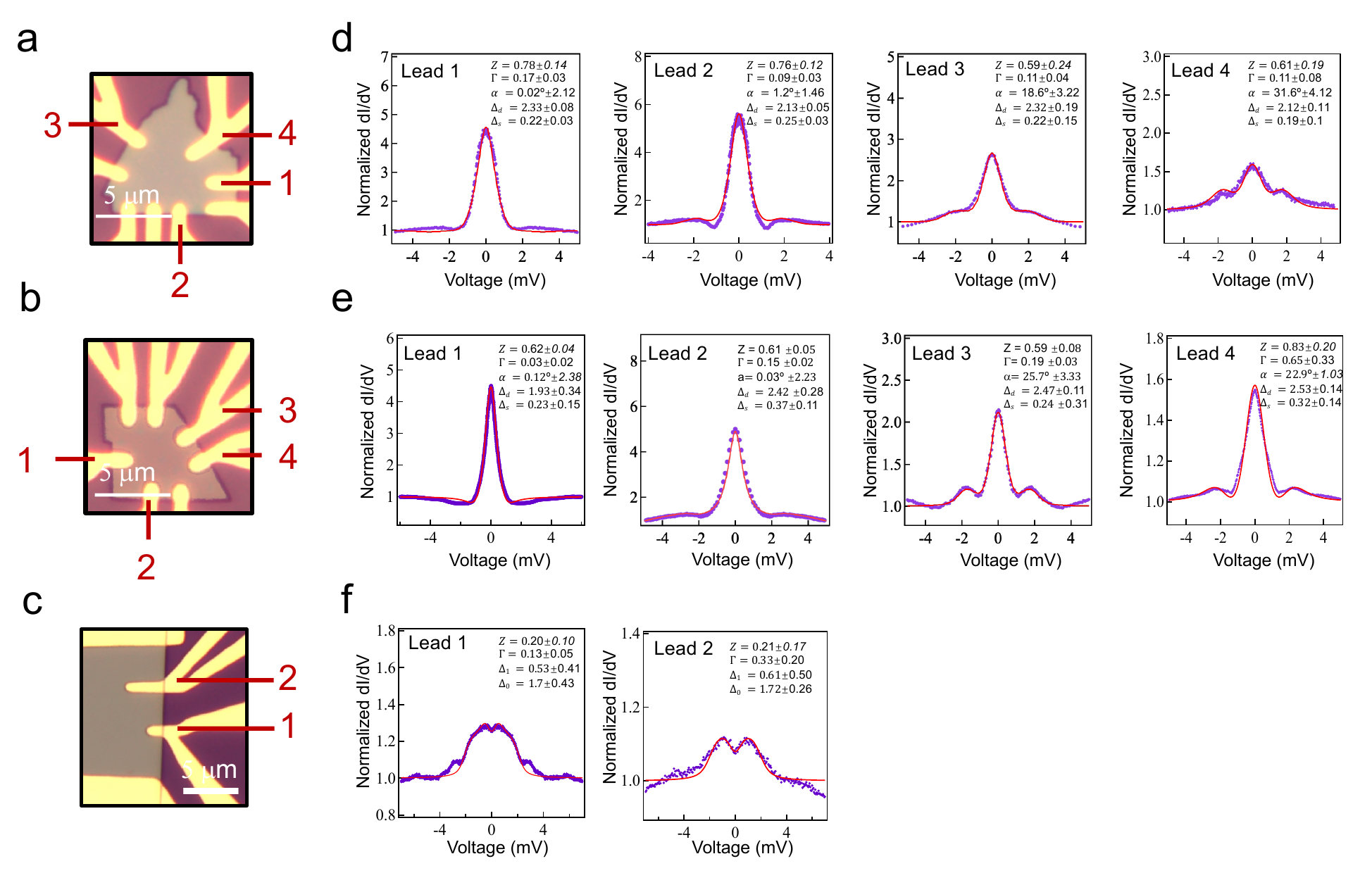}
    \caption{\label{ExtenedFig:BTKfitARS} \textbf{Fit of the normalized low-temperature conductance on FeSe flakes.} \textbf{a} - \textbf{c,} Optical image of the FeSe devices. \textbf{d} - \textbf{f,} The normalized conductance is probed from leads on each device (symbols) at 1.3K. \textbf{d} and \textbf{e,} are fitted by the relevant 2D-BTK model (line) assuming a $s$+$d$-wave gap, $\Delta_{s+d}(\mathbf{E,\alpha}) = \Delta_{s}(E) + \Delta_{d}(E)\cos(2\alpha+\frac{\pi}{2})$ with the parameters listed. \textbf{f,} is using extended $s$-wave gap, $\Delta_{es}(\mathbf{E,\alpha}) = \Delta_{0}(E) + \Delta_{1}(E)\cos(4\alpha+\pi)$. Here, the three-terminal dI/dV spectra in (\textbf{d}) are probed from marked leads from the device (\textbf{a}). (\textbf{e}) and (\textbf{f}) are probed from (\textbf{b}) and (\textbf{c}) respectively. 
    }
\end{extendedfigure}

\newpage
\begin{extendedfigure}[t]
    \centering
    \includegraphics[width=0.5\linewidth]{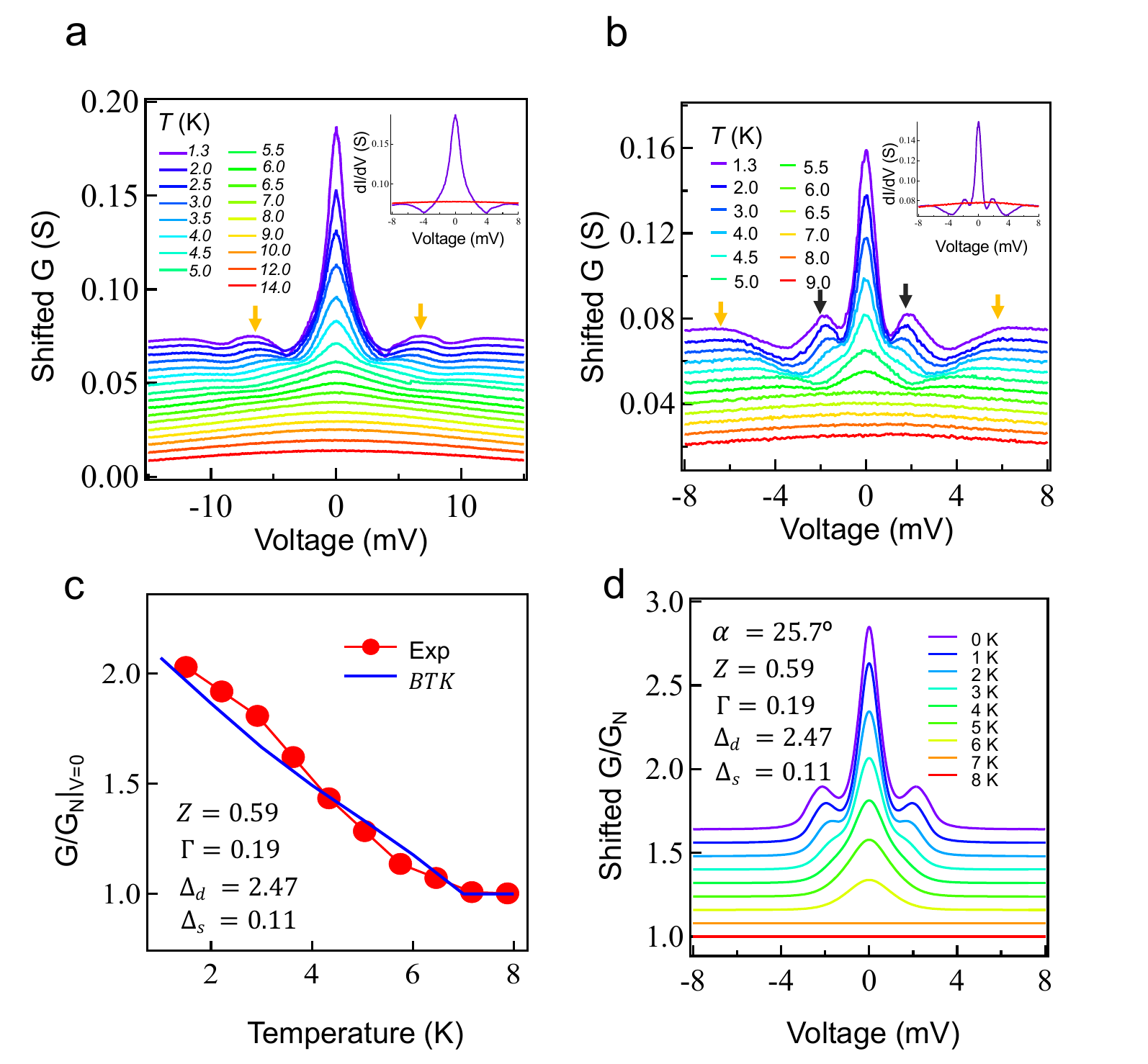}
    \caption{\label{ExtenedFig:TdependentARS} \textbf{Temperature dependence of ABS spectra.} \textbf{a} and \textbf{b,} Raw conductance curves of contact at (\textbf{a}) $\alpha = 0 ^{\circ}$ and (\textbf{b}) $\alpha = 26 ^{\circ}$, starting at 1.3 K (top curve) with increasing temperatures vertically shifted down for clarity. Black arrows indicate the coherence peaks at around 2 meV.  Conductance bumps are also observed near $\pm6$ meV (orange arrows). The inset shows comparison between the T = 1.3 K and T = 14 K without any shift, showing the $G_N$ is temperature independent. The ZBCP and coherence-peak features gradually smear due to thermal broadening and are fully suppressed at $T_{c} \approx 6 ~K$.  \textbf{c,} Zero-bias conductance versus temperature. The experimental (EXP) data is extracted from (\textbf{b}). The theoretical curve (BTK) is calculated with the 2D-BTK model for a $d$-wave sOP. The parameters are obtained by fitting the curve in (\textbf{b}). \textbf{d,} Theoretical simulated spectra of temperature evolution on  $s$+$d$-wave order parameter at 0 to 8 K, calculated with the 2D-BTK model. These theoretical results reproduce the experimental results in (\textbf{b}). 
    }
\end{extendedfigure}

\newpage
\begin{extendedfigure}[t]
    \centering
    \includegraphics[width=0.75\linewidth]{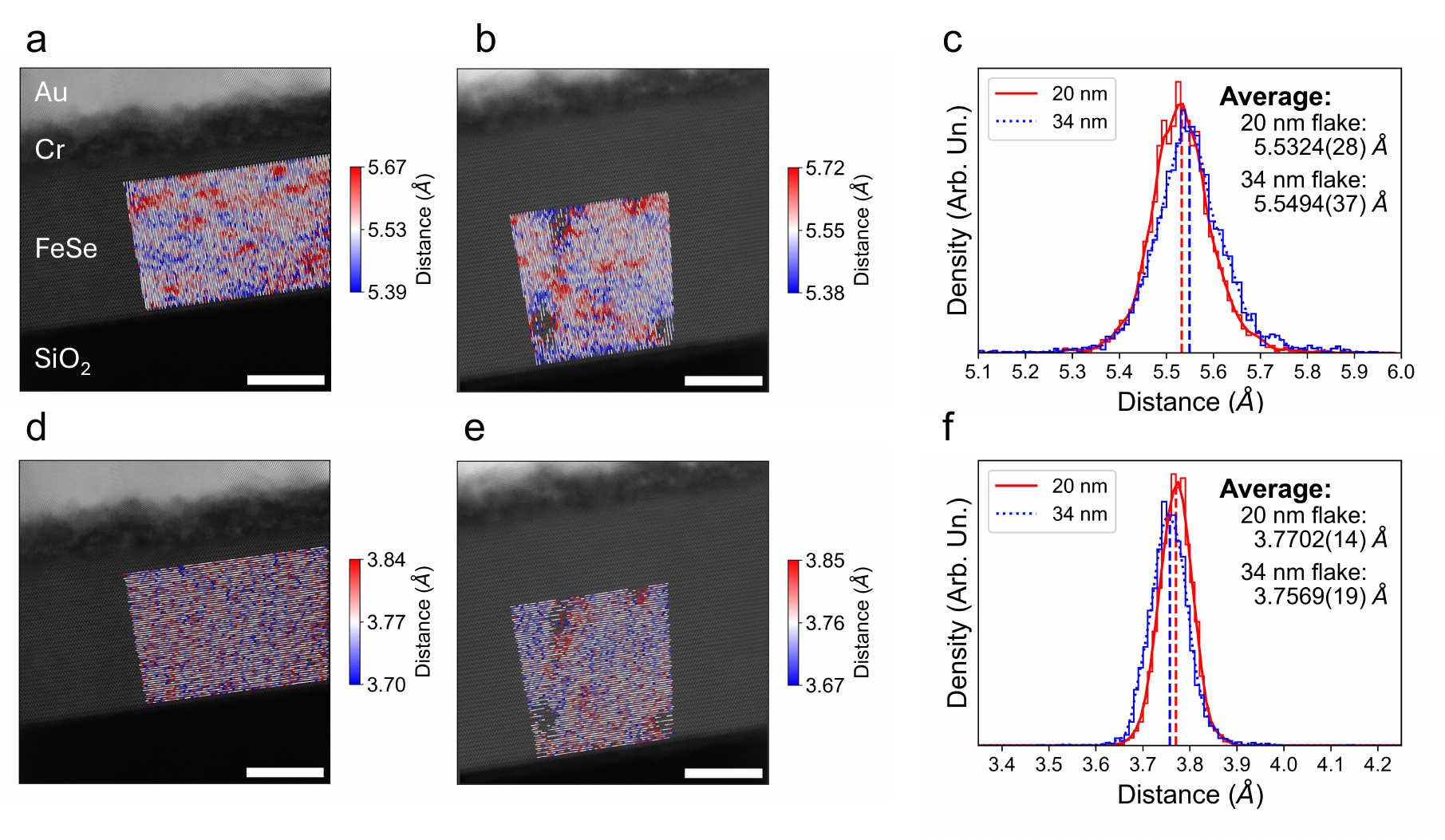}
    \caption{\label{ExtenedFig:STEM-HAADF} \textbf{Lattice parameter measurements of two flake devices from STEM-HAADF images.} \textbf{a,} [001] lattice parameter measured for a 20 nm flake, and \textbf{b,} a 34 nm flake. All measurements are between Se atom columns. \textbf{c,} distribution of measured inter-column distances. The average value is indicated with a 99\% confidence interval. \textbf{d,} - \textbf{f,} same as \textbf{a,} - \textbf{c,} but for the [100] lattice parameter. Only a subsection of each image was selected for measurement to avoid areas with crystalline defects which caused blurring of the atomic columns. Scale bar: 10 nm.
    }
\end{extendedfigure}

\newpage
\begin{extendedfigure}[t]
    \centering
    \includegraphics[width=0.7\linewidth]{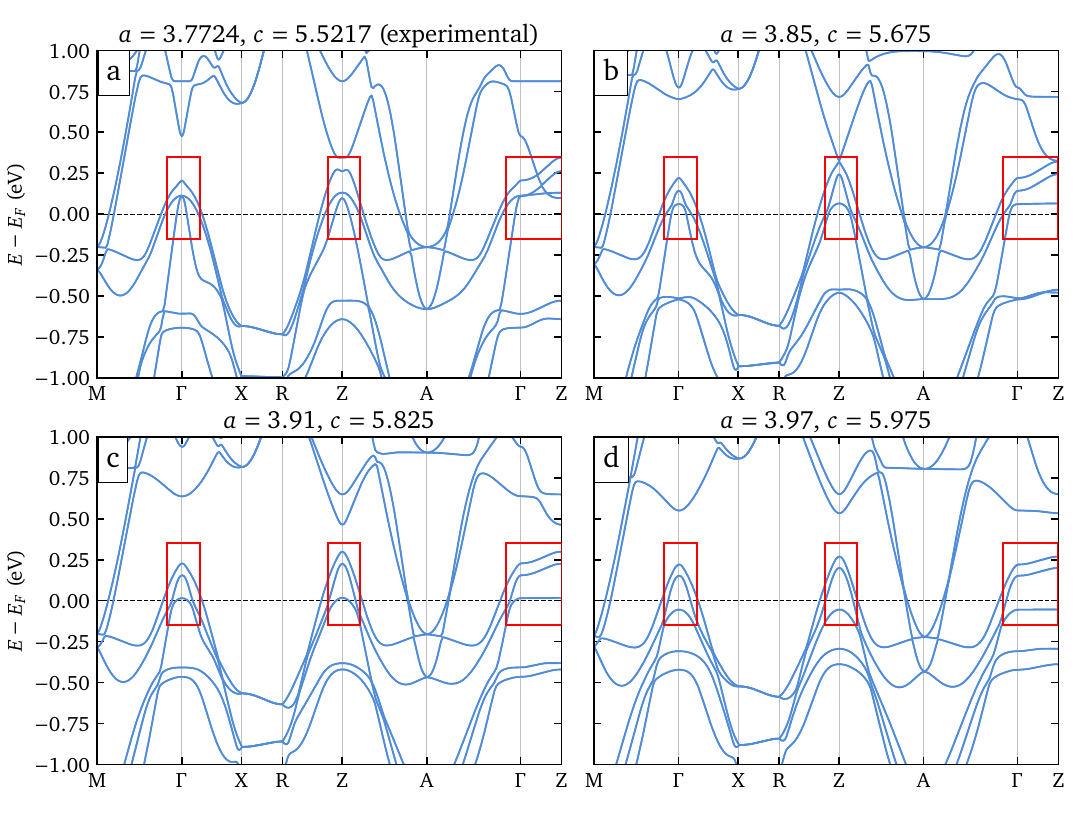}
    \caption{\label{ExtenedFig:DFT} \textbf{Lattice-relaxation-induced band evolution in FeSe} \textbf{a,} Band structure of tetragonal FeSe without lattice relaxation. \textbf{b,} - \textbf{d,} Band structure of FeSe with fixed lattice constant and coupled with custom relaxation for the atomic positions.
    }
\end{extendedfigure}

\newpage
\begin{extendedfigure}[t]
    \centering
    \includegraphics[width=0.65\linewidth]{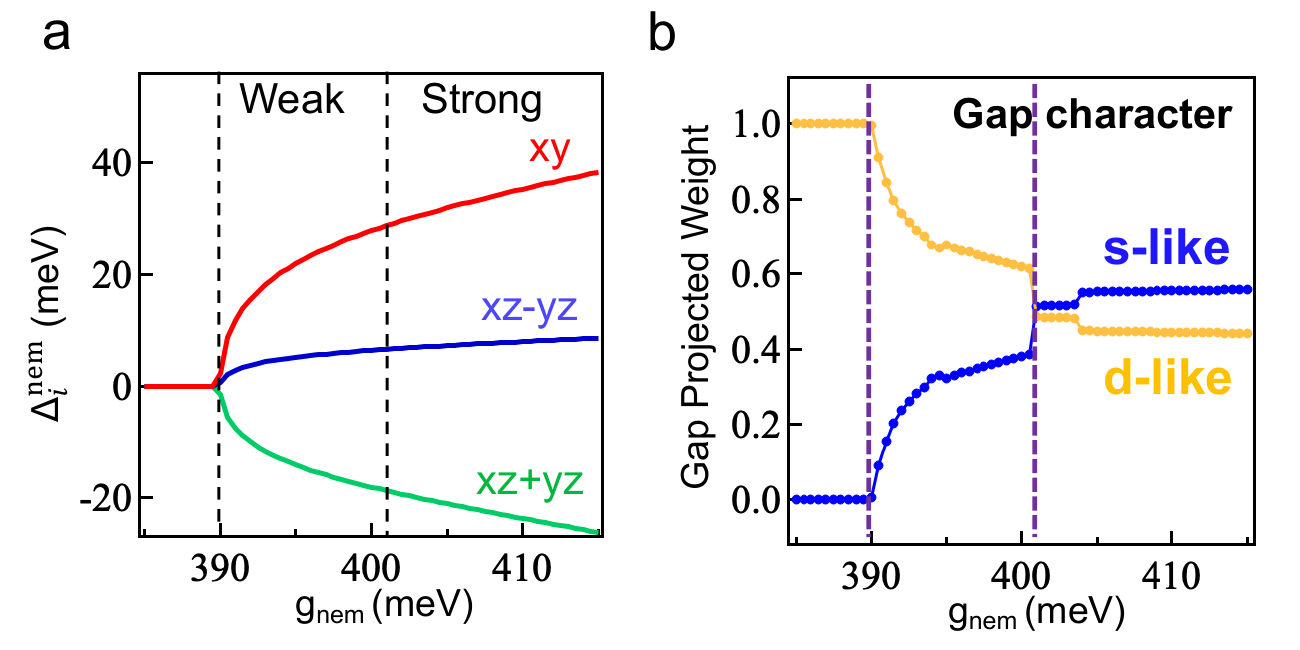}
    \caption{\label{ExtenedFig:RPA} \textbf{Nematicity-controlled superconducting pairing symmetry.} \textbf{a,} Self-consistently calculated magnitude of nematic mean-field order parameters, $|\Delta^{nem}_{\mathrm{i}}|$, as a function of the nematic interaction strength $g_{\mathrm{nem}}$. Interactions are included in the all $B_{1g}$ nematic channels (i = xy, xz-yz or xz+yz) of the unfolded Brillouin zone. \textbf{b,} Evolution of the superconducting gap character with $g_{\mathrm{nem}}$, obtained by solving the linearized superconducting gap equation within the spin/charge-fluctuation RPA framework (see Supplementary Information section X). The gap weights are the projections of the normalized leading gap eigenfunction onto $s$-like and $d$-like pairing channels.
    }
\end{extendedfigure}

\end{document}